\documentclass[12pt]{article}
\usepackage{a4wide,feynarts,amsmath,amssymb,graphicx,psfrag}
\usepackage{colordvi,gnuplotcolors}

\newcommand{\E}[1]{10^{#1}}
\newcommand{\TE}[1]{\cdot 10^{#1}}

\newcommand{\gev}{\text{GeV}}
\newcommand{\tev}{\text{TeV}}
\newcommand{\fb}{\text{fb}}
\newcommand{\pb}{\text{pb}}

\newcommand{\BR}{\text{BR}}

\newcommand{\ma}{ m_{A^0} }

\newcommand{\MSUSY}{ M_{\text{SUSY}} }

\newcommand{\sqrts}{\sqrt{\hat{s}}} 
\newcommand{\MAX}{{\text{max}}}
\newcommand{\MIN}{{\text{min}}}
\newcommand{\LAB}{{\text{lab}}}
\newcommand{\JET}{\text{jet}}
\newcommand{\EFF}{{\text{eff}}}
\newcommand{\MSSM}{{\text{MSSM}}}
\newcommand{\SM}{{\text{SM}}}

\newcommand{\tb}{\tan\beta}

\begin{document}
\setlength{\unitlength}{1mm}

\begin{titlepage}
\begin{flushright}
IPPP/07/18\\
DCPT/07/36\\
MPP-2007-58\\
\end{flushright}
\vspace{2cm}

\begin{center}
{\Large \bf
Distributions for MSSM    
Higgs boson + jet \\[.4cm]   
production at hadron colliders\\
}
\vspace{2.5cm}
{\large \sc Oliver~Brein \footnote{E-mail: obr@mppmu.mpg.de}} 
{\large and} {\large \sc 
Wolfgang~Hollik \footnote{E-mail: hollik@mppmu.mpg.de}
}  \\[0.8cm]
{\normalsize\em 
$^1$ Institute for Particle Physics Phenomenology,\\
University of Durham, DH1 3LE, Durham, United Kingdom\\[.3cm]
$^2$ Max-Planck-Institut f\"ur Physik,\\
        F\"ohringer Ring 6, D-80805 M\"unchen, Germany}\\[1cm]
\end{center}

\begin{abstract}
We present
pseudorapidity and transverse momentum distributions for
the cross section for the production of the lightest 
neutral Higgs boson in association
with a high-$p_T$ hadronic jet, calculated in the framework of 
the minimal supersymmetric standard model (MSSM).
We discuss
the theoretical predictions for the differential
cross sections at the Large Hadron Collider and the Tevatron
for the most common benchmark scenarios.
\end{abstract}

\vspace*{2cm}

\end{titlepage}

\section{Introduction}

The appearance of a light Higgs boson, which might be similar
to that of the Standard Model (SM), 
is a specific feature of the  minimal supersymmetric Standard Model (MSSM),
and distinguishing the various scenarios is an important task
for coming experiments at hadron colliders.
Higgs discovery through the main decay channel into a pair of  
$b\bar b$ quarks will be difficult~\cite{CMS-TDR},
in particular in the mass range below 140 GeV 
which is common to both the SM and the MSSM, and 
experimental investigations have to deal with 
signatures resulting from the more rare decay modes.

Besides inclusive single Higgs production, 
Higgs boson production in association with a high-$p_T$ hadronic jet
provides a useful channel for Higgs searches at hadron colliders,
which allows for refined cuts increasing the signal-to-background ratio.
Specifically, for the SM Higgs boson~\cite{SMHJcalc}, 
simulations considering the decay channels
$H\rightarrow \gamma\gamma$ \cite{ADIKSS,Zmushko} 
and $H\rightarrow \tau^+ \tau^-$ \cite{mellado-etal}
have shown promising signal and background results for
the ATLAS detector.

In the meantime, 
a lot of progress has been made
towards improving the SM predictions.
The fully differential distribution for Higgs production
at next-to-next-to-leading order QCD accuracy
has become available \cite{AMP}, improved by
resummation of logarithmically enhanced  
terms for low $p_T$~\cite{grazzini-etal}.
Higher-order corrections to differential cross sections for 
a Higgs boson associated with a high-$p_T$ jet
have been obtained explicitly: the next-to-leading order QCD corrections 
in the large top-mass limit \cite{kunszt-etal}
and, recently, the corresponding resummation of soft-gluon 
emission effects \cite{kulesza-etal}.

For the analogous MSSM processes,
$p_T$-distributions 
have been studied in the limit of vanishing superpartner contributions 
at leading order~\cite{BField-etal} and were
improved recently by soft-gluon
resummation effects \cite{langenegger-etal}.
This limit is 
usually a good approximation when the superpartners
are heavy, at a mass scale around $1\,\tev$.
The general case was treated in
a previous paper, where we studied the MSSM prediction for the total hadronic
cross section for the production of the lightest neutral Higgs boson $h^0$
in association with a high-$p_T$ jet including all superpartner loop
contributions at leading order~\cite{hjet-own},
thus extending the MSSM predictions also to
the case of light supersymmetric
partners of quarks and gluons.

In the present paper, we complete our study by presenting results 
for the distributions of  
pseudorapidity and transverse momentum, $\eta_3$ and $p_T$.
Section 2 briefly reviews the contributing parton processes, and
Section 3 introduces the  expressions for the differential 
hadronic cross sections 
$d\sigma/d\eta_3$, $d\sigma/d p_T$ and $d^2\sigma/d\eta_3 d p_T$
in the presence of cuts.
In Section~4 we present numerical results for the distributions
at the LHC and the Tevatron, based on
MSSM parameters around
specific benchmark scenarios, in comparison with the SM results.

\section{Partonic processes}

\begin{figure}[tb]
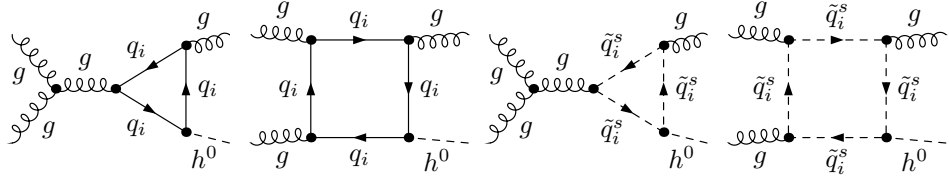

\begin{center}
\begin{footnotesize}
\unitlength=1bp%

\begin{feynartspicture}(432,90)(4,1)
\FADiagram{}
\FAProp(0.,15.)(4.,10.)(0.,){/Cycles}{0}
\FALabel(1.26965,12.0117)[tr]{$g$}
\FAProp(0.,5.)(4.,10.)(0.,){/Cycles}{0}
\FALabel(2.73035,7.01172)[tl]{$g$}
\FAProp(20.,15.)(16.,14.)(0.,){/Cycles}{0}
\FALabel(17.6241,15.5237)[b]{$g$}
\FAProp(20.,5.)(16.,6.)(0.,){/ScalarDash}{0}
\FALabel(17.6847,4.71881)[t]{$h^0$}
\FAProp(4.,10.)(9.5,10.)(0.,){/Cycles}{0}
\FALabel(6.75,11.77)[b]{$g$}
\FAProp(16.,14.)(16.,6.)(0.,){/Straight}{-1}
\FALabel(17.07,10.)[l]{$q_i$}
\FAProp(16.,14.)(9.5,10.)(0.,){/Straight}{1}
\FALabel(12.4176,12.8401)[br]{$q_i$}
\FAProp(16.,6.)(9.5,10.)(0.,){/Straight}{-1}
\FALabel(12.4176,7.15993)[tr]{$q_i$}
\FAVert(4.,10.){0}
\FAVert(16.,14.){0}
\FAVert(16.,6.){0}
\FAVert(9.5,10.){0}

\FADiagram{}
\FAProp(0.,15.)(5.5,14.5)(0.,){/Cycles}{0}
\FALabel(2.95371,16.5108)[b]{$g$}
\FAProp(0.,5.)(5.5,5.5)(0.,){/Cycles}{0}
\FALabel(2.89033,4.18637)[t]{$g$}
\FAProp(20.,15.)(14.5,14.5)(0.,){/Cycles}{0}
\FALabel(17.1097,15.8136)[b]{$g$}
\FAProp(20.,5.)(14.5,5.5)(0.,){/ScalarDash}{0}
\FALabel(17.1323,4.43534)[t]{$h^0$}
\FAProp(5.5,14.5)(5.5,5.5)(0.,){/Straight}{-1}
\FALabel(4.43,10.)[r]{$q_i$}
\FAProp(5.5,14.5)(14.5,14.5)(0.,){/Straight}{1}
\FALabel(10.,15.57)[b]{$q_i$}
\FAProp(5.5,5.5)(14.5,5.5)(0.,){/Straight}{-1}
\FALabel(10.,4.43)[t]{$q_i$}
\FAProp(14.5,14.5)(14.5,5.5)(0.,){/Straight}{1}
\FALabel(15.57,10.)[l]{$q_i$}
\FAVert(5.5,14.5){0}
\FAVert(5.5,5.5){0}
\FAVert(14.5,14.5){0}
\FAVert(14.5,5.5){0}

\FADiagram{}
\FAProp(0.,15.)(4.,10.)(0.,){/Cycles}{0}
\FALabel(1.26965,12.0117)[tr]{$g$}
\FAProp(0.,5.)(4.,10.)(0.,){/Cycles}{0}
\FALabel(2.73035,7.01172)[tl]{$g$}
\FAProp(20.,15.)(16.,14.)(0.,){/Cycles}{0}
\FALabel(17.6241,15.5237)[b]{$g$}
\FAProp(20.,5.)(16.,6.)(0.,){/ScalarDash}{0}
\FALabel(17.6847,4.71881)[t]{$h^0$}
\FAProp(4.,10.)(9.5,10.)(0.,){/Cycles}{0}
\FALabel(6.75,11.77)[b]{$g$}
\FAProp(16.,14.)(16.,6.)(0.,){/ScalarDash}{-1}
\FALabel(17.07,10.)[l]{$\tilde q_i^s$}
\FAProp(16.,14.)(9.5,10.)(0.,){/ScalarDash}{1}
\FALabel(12.4176,12.8401)[br]{$\tilde q_i^s$}
\FAProp(16.,6.)(9.5,10.)(0.,){/ScalarDash}{-1}
\FALabel(12.4176,7.15993)[tr]{$\tilde q_i^s$}
\FAVert(4.,10.){0}
\FAVert(16.,14.){0}
\FAVert(16.,6.){0}
\FAVert(9.5,10.){0} 

\FADiagram{}
\FAProp(0.,15.)(5.5,14.5)(0.,){/Cycles}{0}
\FALabel(2.95371,16.5108)[b]{$g$}
\FAProp(0.,5.)(5.5,5.5)(0.,){/Cycles}{0}
\FALabel(2.89033,4.18637)[t]{$g$}
\FAProp(20.,15.)(14.5,14.5)(0.,){/Cycles}{0}
\FALabel(17.1097,15.8136)[b]{$g$}
\FAProp(20.,5.)(14.5,5.5)(0.,){/ScalarDash}{0}
\FALabel(17.1323,4.43534)[t]{$h^0$}
\FAProp(5.5,14.5)(5.5,5.5)(0.,){/ScalarDash}{-1}
\FALabel(4.43,10.)[r]{$\tilde q_i^s$}
\FAProp(5.5,14.5)(14.5,14.5)(0.,){/ScalarDash}{1}
\FALabel(10.,15.57)[b]{$\tilde q_i^s$}
\FAProp(5.5,5.5)(14.5,5.5)(0.,){/ScalarDash}{-1}
\FALabel(10.,4.43)[t]{$\tilde q_i^s$}
\FAProp(14.5,14.5)(14.5,5.5)(0.,){/ScalarDash}{1}
\FALabel(15.57,10.)[l]{$\tilde q_i^s$}
\FAVert(5.5,14.5){0}
\FAVert(5.5,5.5){0}
\FAVert(14.5,14.5){0}
\FAVert(14.5,5.5){0}

\end{feynartspicture}

\end{footnotesize}

\caption{\label{gghg}
Typical quark and squark 
loop graphs for the process $g g \to g h^0$ in leading order.
Feynman graphs with opposite direction of charge flow are not depicted.}
\end{center}
\end{figure}

\begin{figure}[tb]
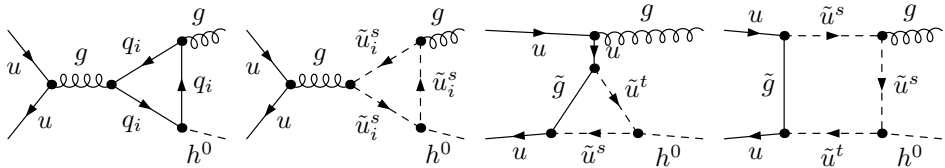

\begin{center}
\begin{footnotesize}
\unitlength=1bp%
\begin{feynartspicture}(432,90)(4,1)
\FADiagram{}
\FAProp(0.,15.)(4.,10.)(0.,){/Straight}{1}
\FALabel(1.26965,12.0117)[tr]{$u$}
\FAProp(0.,5.)(4.,10.)(0.,){/Straight}{-1}
\FALabel(2.73035,7.01172)[tl]{$u$}
\FAProp(20.,15.)(16.,14.)(0.,){/Cycles}{0}
\FALabel(17.6241,15.5237)[b]{$g$}
\FAProp(20.,5.)(16.,6.)(0.,){/ScalarDash}{0}
\FALabel(17.6847,4.71881)[t]{$h^0$}
\FAProp(4.,10.)(9.5,10.)(0.,){/Cycles}{0}
\FALabel(6.75,11.77)[b]{$g$}
\FAProp(16.,14.)(16.,6.)(0.,){/Straight}{-1}
\FALabel(17.07,10.)[l]{$q_i$}
\FAProp(16.,14.)(9.5,10.)(0.,){/Straight}{1}
\FALabel(12.4176,12.8401)[br]{$q_i$}
\FAProp(16.,6.)(9.5,10.)(0.,){/Straight}{-1}
\FALabel(12.4176,7.15993)[tr]{$q_i$}
\FAVert(4.,10.){0}
\FAVert(16.,14.){0}
\FAVert(16.,6.){0}
\FAVert(9.5,10.){0}
 
\FADiagram{}
\FAProp(0.,15.)(4.,10.)(0.,){/Straight}{1}
\FALabel(1.26965,12.0117)[tr]{$u$}
\FAProp(0.,5.)(4.,10.)(0.,){/Straight}{-1}
\FALabel(2.73035,7.01172)[tl]{$u$}
\FAProp(20.,15.)(16.,14.)(0.,){/Cycles}{0}
\FALabel(17.6241,15.5237)[b]{$g$}
\FAProp(20.,5.)(16.,6.)(0.,){/ScalarDash}{0}
\FALabel(17.6847,4.71881)[t]{$h^0$}
\FAProp(4.,10.)(9.5,10.)(0.,){/Cycles}{0}
\FALabel(6.75,11.77)[b]{$g$}
\FAProp(16.,14.)(16.,6.)(0.,){/ScalarDash}{-1}
\FALabel(17.07,10.)[l]{$\tilde u_i^s$}
\FAProp(16.,14.)(9.5,10.)(0.,){/ScalarDash}{1}
\FALabel(12.4176,12.8401)[br]{$\tilde u_i^s$}
\FAProp(16.,6.)(9.5,10.)(0.,){/ScalarDash}{-1}
\FALabel(12.4176,7.15993)[tr]{$\tilde u_i^s$}
\FAVert(4.,10.){0}
\FAVert(16.,14.){0}
\FAVert(16.,6.){0}
\FAVert(9.5,10.){0}

\FADiagram{}
\FAProp(0.,15.)(10.,14.5)(0.,){/Straight}{1}
\FALabel(4.9226,13.6819)[t]{$u$}
\FAProp(0.,5.)(6.,5.5)(0.,){/Straight}{-1}
\FALabel(3.12872,4.18535)[t]{$u$}
\FAProp(20.,15.)(10.,14.5)(0.,){/Cycles}{0}
\FALabel(14.9226,15.8181)[b]{$g$}
\FAProp(20.,5.)(14.,5.5)(0.,){/ScalarDash}{0}
\FALabel(16.892,4.43449)[t]{$h^0$}
\FAProp(10.,14.5)(10.,11.5)(0.,){/Straight}{1}
\FALabel(11.07,13.)[l]{$u$}
\FAProp(6.,5.5)(14.,5.5)(0.,){/ScalarDash}{-1}
\FALabel(10.,4.43)[t]{$\tilde u^s$}
\FAProp(6.,5.5)(10.,11.5)(0.,){/Straight}{0}
\FALabel(7.39833,8.74111)[br]{$\tilde g$}
\FAProp(14.,5.5)(10.,11.5)(0.,){/ScalarDash}{-1}
\FALabel(12.8097,8.87979)[bl]{$\tilde u^t$}
\FAVert(10.,14.5){0}
\FAVert(6.,5.5){0}
\FAVert(14.,5.5){0}
\FAVert(10.,11.5){0}

\FADiagram{}
\FAProp(0.,15.)(5.5,14.5)(0.,){/Straight}{1}
\FALabel(2.89033,15.8136)[b]{$u$}
\FAProp(0.,5.)(5.5,5.5)(0.,){/Straight}{-1}
\FALabel(2.89033,4.18637)[t]{$u$}
\FAProp(20.,15.)(14.5,14.5)(0.,){/Cycles}{0}
\FALabel(17.1097,15.8136)[b]{$g$}
\FAProp(20.,5.)(14.5,5.5)(0.,){/ScalarDash}{0}
\FALabel(17.1323,4.43534)[t]{$h^0$}
\FAProp(5.5,14.5)(5.5,5.5)(0.,){/Straight}{0}
\FALabel(4.68,10.)[r]{$\tilde g$}
\FAProp(5.5,14.5)(14.5,14.5)(0.,){/ScalarDash}{1}
\FALabel(10.,15.57)[b]{$\tilde u^s$}
\FAProp(5.5,5.5)(14.5,5.5)(0.,){/ScalarDash}{-1}
\FALabel(10.,4.43)[t]{$\tilde u^t$}
\FAProp(14.5,14.5)(14.5,5.5)(0.,){/ScalarDash}{1}
\FALabel(15.57,10.)[l]{$\tilde u^s$}
\FAVert(5.5,14.5){0}
\FAVert(5.5,5.5){0}
\FAVert(14.5,14.5){0}
\FAVert(14.5,5.5){0}
\end{feynartspicture}
\end{footnotesize} 
\caption{\label{uuhg}
Typical Feynman graphs for the process $u \bar u \to g h^0$ in
leading order. 
For the scattering of the other quarks ($d,c,s$) the Feynman graphs 
look similar.
}
\end{center}
\end{figure}

\begin{figure}[tb]
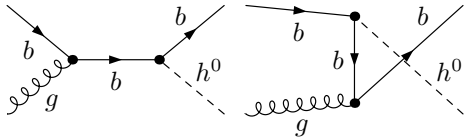

\begin{center}
\begin{footnotesize}
\unitlength=1bp%

\begin{feynartspicture}(432,90)(2,1)

\FADiagram{}
\FAProp(0.,15.)(6.,10.)(0.,){/Straight}{1}
\FALabel(2.48771,11.7893)[tr]{$b$}
\FAProp(0.,5.)(6.,10.)(0.,){/Cycles}{0}
\FALabel(3.51229,6.78926)[tl]{$g$}
\FAProp(20.,15.)(14.,10.)(0.,){/Straight}{-1}
\FALabel(16.4877,13.2107)[br]{$b$}
\FAProp(20.,5.)(14.,10.)(0.,){/ScalarDash}{0}
\FALabel(17.3522,8.01869)[bl]{$h^0$}
\FAProp(6.,10.)(14.,10.)(0.,){/Straight}{1}
\FALabel(10.,8.93)[t]{$b$}
\FAVert(6.,10.){0}
\FAVert(14.,10.){0}

\FADiagram{}
\FAProp(0.,15.)(10.,14.)(0.,){/Straight}{1}
\FALabel(4.84577,13.4377)[t]{$b$}
\FAProp(0.,5.)(10.,6.)(0.,){/Cycles}{0}
\FALabel(5.15423,4.43769)[t]{$g$}
\FAProp(20.,15.)(10.,6.)(0.,){/Straight}{-1}
\FALabel(16.8128,13.2058)[br]{$b$}
\FAProp(20.,5.)(10.,14.)(0.,){/ScalarDash}{0}
\FALabel(17.52,8.02)[bl]{$h^0$}
\FAProp(10.,14.)(10.,6.)(0.,){/Straight}{1}
\FALabel(9.03,10.)[r]{$b$}
\FAVert(10.,14.){0}
\FAVert(10.,6.){0}
\end{feynartspicture}

\end{footnotesize}

\caption{\label{bghb-born}
Feynman graphs for the $b$-quark processes in leading order. 
The graphs represent
the amplitude for the process $b g \to b h^0$ if the time axis
points to the right and $b\bar b \to g h^0$  if 
it
points down.
}
\end{center}
\end{figure}

At the partonic level, production of a Higgs-boson $h^0$ together
with a jet is described generically by 
\begin{align*}
P_1 (k_1) + P_2 (k_2) & 
        \to P_3 (k_3) + h^0 (k_4)
\,,
\end{align*}
involving partons $P_i$ with momenta $k_i = (E_i, \vec k_i)$. 
There are three classes of parton processes:
gluon fusion $g+g\to g+h^0$ (Figure~\ref{gghg}),
quark--gluon scattering $q+g\to q+h^0$ (Figure~\ref{ughu}), 
and 
quark--anti-quark annihilation $q +\bar q \to g+ h^0$ (Figure~\ref{uuhg}). 
While gluon fusion is an entirely loop-induced process, 
the other two classes also 
get contributions from tree-level $b$-quark initiated processes 
(see Figure \ref{bghb-born}).
Those Born-type processes are in general 
dominant for $m_A \lesssim 120\,\gev$,
while for large values of $m_A$ the loop-induced processes dominate. 
This behaviour is  essentially a consequence of the 
Yukawa-coupling of the lightest MSSM Higgs boson
to $b$-quarks, which can be enhanced for low values of the $A$-boson mass
$m_A$.
Formulae for the relevant couplings occurring in this process
in our notation can be found in \cite{hjet-own,gghphm-and-hw}.

The virtual presence of the 
superpartners in the loop contributions 
modify in addition the overall production rates for 
supersymmetric Higgs bosons. Moreover, they affect 
the angular distributions and thus, 
at the level of hadronic processes,
change rapidity and transverse-momentum distributions
of the Higgs bosons or the jets, respectively.

In terms of the invariant kinematic variables
\begin{align}
\hat{s} = (k_1+k_2)^2, \quad \hat{t} = (k_1-k_3)^2~, 
\end{align}
the partonic differential cross section  
reads
\begin{align}
\frac{d\hat{\sigma}}{d \hat{t}} (\hat s,\hat t)=
 \frac {1}{16 \pi^2 \hat{s}^2} \, 
 \overline{| {\cal M}(\hat{s},\hat{t})|^2} \, ,
\end{align}
with the spin- and colour-summed/averaged $S$-matrix element squared.

Instead of $\hat{t}$, the cross section can be expressed in terms of
other variables, such as the transverse momentum $p_T = k_{3,T}$,
or the scattering angle $\hat{\theta}$ or  rapidity $\hat{y}_3$
in the parton center-of-mass system (cms), yielding the representations
\begin{eqnarray}
& & \frac{d\hat{\sigma}}{d {p}_T}(\hat{s},p_T)  = 
	\frac{d\hat\sigma^{F}}{d {p}_T}(\hat{s},p_T)
	+ \frac{d\hat\sigma^{B}}{d {p}_T}(\hat{s},p_T)
\,,\\
& & \frac{d\hat\sigma^{F[B]}}{d {p}_T}(\hat{s},p_T)  = 
	\frac{2 |\vec k_1|\, p_T}{\sqrt{|\vec k_3|^2 - p_T^{\; 2}}}
	\left.
	\frac{d\hat\sigma}{d\hat t} (\hat s,\hat t)
	\right|_{\hat t = \hat t_{F[B]}(\hat s,p_T)}
\,,\\
& & \frac{d\hat{\sigma}}{d \hat{\theta}}(\hat{s},\hat{\theta})  = 
        2 |\vec k_1| |\vec k_3| \sin\hat{\theta} \,
	\left.
        \frac{d\hat\sigma}{d\hat t} (\hat s,\hat t)
        \right|_{\hat t = \hat t (\hat s, \hat{\theta})}
\,,\\
& & \frac{d\hat{\sigma}}{d \hat{y}_3}(\hat{s},\hat{y}_3)  = 
	2 |\vec k_1| E_3 (1 - \tanh^2\hat y_3)
	\left.
	\frac{d\hat\sigma}{d\hat t} (\hat s,\hat t)
	\right|_{\hat t = \hat t (\hat s, \hat y_3)}
\,,
\end{eqnarray}
with
\begin{eqnarray}
\label{pt-tFB}
& & \hat t_{F [B]}(\hat s, p_T)  = 
 -2 E_1 E_3 +[-] \;
	2 |\vec k_1| \sqrt{|\vec{k_3}|^2 - p_T^{\; 2}}
\,,\\
& & \hat t(\hat s, \hat y_3)  = 
	- 2 E_1 E_3 + 2 |\vec k_1| E_3 \tanh\hat y_3
\,,\\
& & \hat t(\hat s, \hat{\theta})  = 
	- 2 E_1 E_3 + 2 |\vec k_1| |\vec k_3| \cos\hat{\theta}
\,,  \\
& & E_1 = |\vec k_1| = \frac{\sqrt{\hat s}}{2},  \quad
    E_3 = |\vec k_3| = \frac{\hat s - m_h^2}{2 \sqrt{\hat s}} \, .
\end{eqnarray}

\section{Hadronic observables}

Hadron colliders like the LHC (Tevatron) collide protons with protons
(anti-protons) in the center-of-mass system (the laboratory frame)
with a total energy $\sqrt S$
and individual momenta $\vec P$ and $-\vec P$, respectively.
Hadronic cross section are obtained via convolution of the
parton-level cross sections with the parton distributions and
summation over the various contributing partons.
Experimental restrictions to the detectability
of the produced particles are conventionally realized by
imposing specific cuts to the kinematically allowed phase space.
Typically, cuts are imposed on the final-state
transverse momentum and/or the pseudorapidity in order 
to have high-$p_T$ jets not too close to the beam axis.
In our case, we choose the following selection criteria,
\begin{align}
\label{the-cuts}
p_T \equiv |\vec k_3|\sin\hat\theta & > p_T^\MIN\,, & 
| \eta_{3} | & < \eta_{\text{max}}\,,
\end{align}
where $p_T$ and $\eta_{3}$
denote transverse momentum  and pseudorapidity
of the final state parton.
The first condition leads to an energy cut on the invariant mass
of the parton system,
\begin{align}
\label{s-cut}
\sqrts & > p_T^\MIN + \sqrt{m_h^2+(p_T^\MIN)^2} 
        \equiv \sqrt{\hat{s}_0}~,
\end{align}
and to a cut on the scattering angle $\hat\theta$
in the parton cms,
\begin{align}
\sin\hat{\theta} & > \frac{2\, \hat s\, p_T^\MIN}{\hat{s}-m_h^2}
\,. 
\end{align}
The second condition in (\ref{the-cuts})
avoids partons with angles $\theta_{\text{lab}}$
unobservably close to the
beam axis in the laboratory frame.
For an outgoing massless gluon or quark
with momentum $k_3$, $\eta_3$ coincides 
with the parton rapidity $y_{3}^{\text{lab}}$
in the laboratory frame,
which is related to the rapidity $\hat y_3$
in the parton cms via 
$\hat y_3 = y_3^\LAB + \psi(\tau,x)$,
where $\psi$ is determined by the relative velocity 
\begin{align}
\label{def-psi}
\beta = (x^2 - \tau)/(x^2 + \tau) \equiv - \tanh \psi(\tau,x)
\end{align}
of the parton cms with respect to the laboratory frame. 
Thereby, it is assumed that the initial state partons carry 
momentum fractions $x$ and $\tau/x$ of the hadrons they originate from.
The condition $|\eta_3| < \eta_{\rm max}$
can be written as follows,
\begin{align}
\label{ang-cut2}
\arccos(\tanh \eta_{\MAX})& < \theta_{\text{lab}}  
        < \pi - \arccos(\tanh \eta_{\MAX})
\,.
\end{align}
This translates into the following additional cut on the scattering angle in the 
parton cms,
\begin{align}
\hat\theta_\MIN & \leq \hat\theta \leq \hat\theta_\MAX
\end{align}
with
\begin{align}
\cos\hat\theta_\MIN & =
	\frac{(\tau-x^2)+(\tau+x^2)\tanh\eta_\MAX}{(\tau+x^2)+(\tau-x^2)\tanh\eta_\MAX}
\,,\\
\cos\hat\theta_\MAX & =
	\frac{(\tau-x^2)-(\tau+x^2)\tanh\eta_\MAX}{(\tau+x^2)-(\tau-x^2)\tanh\eta_\MAX}
\,.
\end{align}

Thus, in the additional presence of the pseudorapidity cut
the partonic cross section for initial state partons $n$ and $m$, with momenta
$x\vec P$ and $-(\tau/x)\vec P$,
depends on $p_T^\MIN$, $\eta_\MAX$
and both variables $\tau$ and $x$,
\begin{align}
\label{parton-xs}
\hat\sigma_{nm}(\tau,x,\eta_\MAX,p_T^\MIN) 
	& = \int_{\hat\theta_\MIN}^{\hat\theta_\MAX} 
	\!d\hat\theta\;
	\left.
	\frac{d\hat\sigma_{nm}}{d\hat\theta}
         (\hat s, \hat\theta) \;
        \Theta(p_T(\hat s, \hat\theta) - p_T^\MIN) 
	\right|_{\hat s=\tau S}
\,.
\end{align}

\subsection{Integrated cross section} 

The hadronic cross section for $h^0$ + jet production 
from two colliding hadrons, $A$ and $B$,
contains the parton densities and the  
partonic cross sections 
$\hat\sigma_{n m}$ from (\ref{parton-xs})
as follows~\cite{pQCD-lect},
\begin{multline}\label{hadronx}
\sigma_{A B}(S,\eta_\MAX, p_T^\MIN) =  
\sum_{\{n,m\}}^{} 
\int_{\tau_0}^{1} d\tau \int_{\tau}^{1} \!\! dx\,
        \frac{1}{(1+\delta_{nm})\, x} \times\\
\bigg\{  f_{n/A} (x,\mu_F) f_{m/B} (\frac{\tau}{x},\mu_F) \,
        \hat \sigma_{n m}
                (\tau, x, \eta_\MAX, p_T^\MIN)\\
+ f_{m/A} (x,\mu_F) f_{n/B} (\frac{\tau}{x},\mu_F) \,
        \hat \sigma_{m n}        
                (\tau, x, \eta_\MAX, p_T^\MIN)
\bigg\},
\end{multline}
where $f_{n/A} (x,\mu_F)$ denotes the density of partons of type $n$
in the hadron $A$ carrying a fraction $x$ of the hadron momentum
at the factorization scale $\mu_F$ 
[$(A,B)=(p,p)$ for the LHC and $(p,\bar{p})$
for the Tevatron].
The sum over unordered pairs of incoming partons runs over 
$\{n,m\} = gg, qg, q\bar q$ with the outgoing parton
$g, q, g$ respectively, for the various channels.
The lower bound of the $\tau$-integration ($\tau_0$) is determined by
the minimal invariant mass of the parton system, $\hat s_0 = \tau_0 S$,
according to~(\ref{s-cut}).

Care has to be taken for the fact that a forward-scattered
parton $n$ out of hadron $A$ combines with a backward-scattered parton $n$
out of hadron $B$ or vice-versa to a given value of $\theta_\LAB$ 
or $\eta_3$.
Hence, a distinction has to be made 
in the notation for the integrated partonic cross section:
$\hat\sigma_{nm}$ has the parton $n$ moving in the direction of 
the incoming hadron $A$ in the laboratory frame and $\hat\sigma_{mn}$
has it in the opposite direction.

\subsection{Transverse-momentum distribution}

The hadronic cross section differential in $p_T$ is given
by the convolution integral of the corresponding 
partonic differential cross section\footnote{In the following we drop
the factorization scale $\mu_F$ in the notation.},
\begin{multline}
\label{ds-dpt}
\frac{d\sigma_{AB}} {d p_T} (S, p_T,  \eta_\MAX)
     = 
\sum_{\{n,m\}}^{}
\int_{\tau_0}^{1} \!\! d\tau
\int_{\tau}^{1} \!\! \frac{dx}{x}\,
\bigg\{
\frac{f_{n/A} (x) f_{m/B}(\frac{\tau}{x})}{1+\delta_{nm}}\times\\
\bigg[
\frac{d\hat\sigma^F_{n m}}{d p_T}
(\tau S, p_T) \; \Theta(\eta_\MAX - |\eta^{nm}_{3,F}(\tau,x,p_T)|)
 \bigg]
+ \bigg[ F \leftrightarrow B \bigg]\bigg\}
 + \; \bigg\{m \leftrightarrow n \bigg\}
\,,
\end{multline}
with
\begin{align*}
\eta^{nm}_{3,F[B]}(\tau,x,p_T) 
	& = - \psi(\tau,x)
+[-]{\rm arcosh}\, \frac{\tau S - m_h^2}{2 p_T \sqrt{\tau S} }
\,, & 
\eta^{mn}_{3,F} & = \eta^{nm}_{3,B}\,, & \eta^{mn}_{3,B} & = \eta^{nm}_{3,F} \,,
\end{align*}
and $\psi$ from (\ref{def-psi}).

\subsection{Pseudorapidity distribution}

Taking into account the two directions 
of motion of the initial state partons
in the laboratory system it is convenient to define 
\begin{align}
\hat y_3^{nm} (\eta_3,\tau,x) & = \eta_3 + \psi(\tau,x)\,, &
\hat y_3^{mn} (\eta_3,\tau,x) & = - \hat y_3^{nm} (\eta_3,\tau,x)
\,.
\end{align}
With this convention
the formula for the hadronic cross section 
differential in $\eta_3$
reads as follows,
\begin{multline}
\label{ds-deta3}
\frac{d\sigma_{AB}} {d \eta_3} (S, \eta_3, p_T^\MIN) =
\sum_{\{n,m\}}^{}
\int_{\tau_0}^{1} \!\! d\tau
\int_{\tau}^{1} \!\! \frac{dx}{x}\,
\bigg\{\\
\frac{f_{n/A} (x) f_{m/B}(\frac{\tau}{x})}{1+\delta_{nm}}  \;
\frac{d\hat\sigma_{n m}} {d \hat y_3} (\tau S, \hat y_3)
\bigg|_{\hat y_3=\hat y_3^{nm}}
 \; \Theta(p_T(\tau, x, \eta_3)-p_T^\MIN)
+\; (n \leftrightarrow m)
\bigg\}
\, 
\end{multline}
where
\begin{align}
\label{pT}
p_T(\tau, x, \eta_3) & =
\frac{\tau S - m_h^2}{2 \sqrt{\tau S} \, 
\cosh( \eta_3 +\psi(x,\tau) ) } \, .
\end{align}

\subsection{Two-fold differential cross section}

The hadronic cross section differential in $p_T$ and $\eta_3$ 
can be written as 
an integral over a single parameter, which can be chosen
to be the parton momentum fraction $x$.
The expression follows from the representation
\begin{eqnarray}
\label{d2s}
& & \frac{d^2\sigma_{AB}}{d p_T d\eta_3} (S, \eta_3, p_T) =  \\
& & \sum_{\{n,m\}}^{}
\int_{\tau_0}^{1} \! d\tau \,
\int_{0}^{1} \!\! \frac{dx}{x}\, \Theta(x-\tau)\,
           G_{nm}(x,\tau,\eta_3)\; 
         \delta \! \left( p_T(\tau,x,\eta_3) - p_T) \right) 
         \, ,  \nonumber
\end{eqnarray}
with 
\begin{eqnarray}
& & G_{nm} (x,\tau,\eta_3)  =
\frac{f_{n/A} (x) f_{m/B}(\frac{\tau}{x})}{1+\delta_{nm}}
\;\frac{d\hat\sigma_{n m}}{d \hat y_3} (\tau S, \hat y_3)
\bigg|_{\hat y_3=\hat y_3^{nm}(\eta_3,\tau,x)}
+(n \leftrightarrow m)\, ,
\end{eqnarray}
by performing the $\tau$-integration with the help of 
the $\delta$-function, yielding
\begin{eqnarray}
\label{d2s-dy-dpt}
& & \frac{d^2\sigma_{AB}}{d p_T d\eta_3} (S, \eta_3, p_T) =  \\
& & 
\sum_{\{n,m\}}^{}
\int_{0}^{1} \!\! \frac{dx}{x}\, 
\Bigg[
\Theta(x-\tau)\Theta(\tau-\tau_0)
\frac{x}{\sqrt{S}}
\frac{x^2 S e^{-\eta_3} + m_h^2 e^{\eta_3}}
{(x \sqrt{S} - p_T e^{\eta_3})^2} \, G_{nm}(x,\tau,\eta_3)
\bigg]_{\tau = \tau_1(x,\eta_3,p_T)}
\nonumber
\end{eqnarray}
where
\begin{eqnarray}
& & \tau_1(x,\eta_3,p_T)  = x \left(
\frac{x\, p_T\, e^{-\eta_3} +\frac{m_h^2}{\sqrt{S}}}
{x \sqrt{S} -p_T\, e^{\eta_3}}
\right)
\end{eqnarray}
fulfils the relation
\begin{eqnarray}
& & p_T(x,\tau_1,\eta_3) - p_T    = 0 
\end{eqnarray}
with  $p_T(x,\tau,\eta_3)$ from (\ref{pT}).

\section{Numerical results}

In the following discussion we want to illustrate
the MSSM predictions for the 
pseu\-do\-ra\-pi\-di\-ty and transverse momentum distributions of the
hadronic processes $pp \to h^0 + \text{jet} + X$ 
and $p\bar p \to h^0 + \text{jet} + X$
and outline differences between MSSM and SM predictions.
For comparison of a given MSSM scenario 
with the SM we choose the SM Higgs mass
to have the same value as the $h^0$ boson  
in that MSSM scenario.
For the numerical evaluation we use the cuts (\ref{the-cuts})
with $p_T^\MIN = 30\,\gev$ 
and $\eta_{\text{max}} = 4.5$ as standard cuts,
which have been used in previous Standard Model
studies for the LHC~\cite{ADIKSS,Zmushko}.

The evaluation has been carried out with the MRST 
parton distribution functions~\cite{MRST}, with the renormalisation
scale $\mu_R$ and factorisation scale for the gluon and the light quarks
$\mu_F^{(g)}$ chosen both equal to $m_h$. 
For the bottom-quark factorisation scale $\mu_F^{(b)}$
we choose $m_h/4$, inspired by the NNLO prediction for the
process $b\bar b \to h^0$ \cite{bbh-NNLO} where it proved to
be the proper scale choice, as anticipated by several
authors \cite{bscale-voodoo}.
For the strong coupling constant $\alpha_S(\mu_R)$, we use the formula 
including the two-loop QCD corrections (see e.g. Ref.~\cite{pdg2006})
for $n_f=5$ with $\Lambda^5_{QCD}=204.8$ MeV.

\subsection{Parameters}\label{parameters}

We adopt for our discussion the MSSM 
benchmark scenarios
for the Higgs search at LEP
suggested in \cite{improvedbm} 
except for the large-$\mu$ scenario which has been
ruled out by LEP data \cite{LEP-susy-Higgs}.
We are interested in effects from the virtual superpartners.
Therefore, the no-mixing and $m_h^\MAX$ scenario
are generalised to have a 
lower common sfermion mass scale $\MSUSY$ 
than in the original proposal. 
The two remaining scenarios in \cite{improvedbm}, 
the  gluophobic and small-$\alpha_{\text{eff}}$ scenario,
have already a rather low sfermion mass scale.

Although our aim here is mainly exemplary we try to take into account 
relevant parameter constraints from previous experiments.

Firstly, we calculate for each parameter point the MSSM predictions for 
$m_{h^0}$ and $\sigma(e^+ e^- \to h^0 Z)\times\BR(h^0\to b\bar b)$
and exclude it if the $m_{h^0}$-dependent LEP-bound on $\sigma\times\BR$
is violated (according to Table 14(b) of \cite{LEP-susy-Higgs}).
We use FeynHiggs 2.5.1 \cite{FH} for calculating the $m_{h^0}$ prediction
and allow for a theoretical uncertainty of $3 \,\gev$.

Secondly, we calculate the leading order MSSM prediction for the 
branching ratio $\BR(B \to X_s \gamma)$ \cite{bsg-mssm} and exclude parameter points
if the prediction falls outside of the range $(3.55 \pm 1.71)\TE{-4}$.
This range is determined by using the experimental central value
\cite{bsg-exp} and 
adding up the experimental $3\sigma$ interval ($\approx \E{-4}$) 
and an estimate of the independent theoretical uncertainty ($0.71\TE{-4}$). 
The latter estimate (20\%) is guided by the 
detailed discussion of theoretical uncertainties for the SM prediction
\cite{misiak-etal}.

Furthermore, we took care that all mass exclusion limits from direct search results
for superpartner particles \cite{pdg2006} are met and that the dominant 
squark-contribution to the electroweak $\rho$-parameter\cite{rho-param} 
stay within $\pm 0.0025$.

Interestingly, applying the rather conservative bound derived from the
$\BR(B \to X_s \gamma)$ prediction, it turns out that
the gluophobic scenario is ruled out in the range we study,
$\{m_A \in [50\,\gev ,1000\,\gev]$, $\tb \in [1,50]\}$.
For this scenario $\BR(B \to X_s \gamma)$ is notoriously too large
with typical values of the order of $\E{-3}$ and 
the lowest value being $7\TE{-4}$.
We do not include results for this scenario here.

A brief specification of the scenarios investigated is following.
\begin{description}
\item{{\em no-mixing$(700)$ scenario }:} 
The soft-breaking sfermion mass parameter is
set to $\MSUSY = 700\,\gev$. For $\MSUSY$ significantly below $700\,\gev$
the whole range of the $m_A$-$\tb$-plane which we study is ruled out. 
The off-diagonal term $X_t$ ($= A_t-\mu\cot\beta$) 
in the top-squark mass matrix is zero,
corresponding to a local minimum of $m_h$ as a function of $X_t$.
The supersymmetric Higgsino mass parameter $\mu$ is set to $- 200 \,\gev$,
the gaugino mass parameters to $M_1 = M_2 = 200 \,\gev$,
and the gluino mass to
$M_{\tilde g} = 800 \,\gev$.
When $\tb$ is changed, $A_t$ is changed accordingly to insure $X_t = 0$. 
The settings of the other soft-breaking scalar-quark Higgs couplings 
are $A_b = A_t$ and $A_q = 0$ ($q=u,d,c,s$).

The input for the
Higgs sector is specified by $m_A = 500\,\gev$ and $\tb=35$.

\item{{\em $m_h^\MAX(400)$ scenario }:}
$X_t$ is set to $2 \MSUSY$
which yields the maximal value of $m_h$ with respect to stop mixing
effects.
We set $\MSUSY=400\,\gev$ and
the other parameters are chosen as in the previous scenario.

In the $m_h^\MAX(400)$ scenario small values of $m_A$ are still allowed.
Hence we examine two Higgs sector scenarios: $m_A = 110\,\gev$, $\tb=30$,
and $m_A = 400\,\gev$, $\tb=30$.
The former leads to the dominance of $b$-quark initiated processes,
while the latter is dominated by the 
loop-induced processes~\cite{hjet-own}.

\item{{\em small-$\alpha_{\text{eff}}$ scenario} :}
This scenario gives rise to suppressed branching ratios for the 
decays $h^0\to b\bar b$ 
and $\tau^+ \tau^-$, especially
for large $\tb$ and moderate values of $m_A$.
The settings are: $\MSUSY=800\,\gev$,
$X_t=-1100\,\gev$,
$M_1=M_2=500\,\gev$, $\mu=2000\,\gev$
and $M_{\tilde g} =500\,\gev$.

We choose $m_A = 400\,\gev$ and $\tb=30$ in the Higgs sector.
Values $m_A \lesssim 300$ are already ruled out, and therefore all 
scenarios with dominance of the $b$-quark initiated processes as well.
\end{description}

\subsection{Differential cross sections at the LHC}

The crucial parameter determining the properties of $h^0$+jet production
in the MSSM is $m_A$ \cite{hjet-own}. 
For $m_A \lesssim 120\,\gev$ and $\tb$ not too small ($\gtrsim 5$) 
the $b$-quark initiated processes (see Figure~\ref{bghb-born}) 
dominate the production rate by far, 
while for larger $m_A$ this role is taken over by the loop-induced
processes (see Figs.~\ref{gghg} to \ref{uuhg}).
Accordingly, we split our discussion 
into the high-$m_A$ and low-$m_A$ cases.

\subsubsection{High {\boldmath{$m_A$}}}

The influence of 
rather light, yet not excluded, superpartners 
on the total hadronic cross section
has been demonstrated to be strong \cite{hjet-own}. 
In particular for the $m_h^\MAX(400)$ scenario with 
$\MSUSY=400\,\gev$
the MSSM cross section for $m_A > 200\,\gev$ and any $\tb \in [1,50]$
is reduced by about $20 - 40\%$ compared to the SM.
Here, we investigate the impact on the shape of the
differential distributions
with respect to the SM.
\smallskip

\begin{figure}[ht]
\psfrag{DELTA}{\LARGE $\delta\; [\%]$}
\psfrag{DSIGMADY}[t][b]{\LARGE $d\sigma/d\eta_3\; [\pb]$}
\psfrag{DSIGMADPT}[t][b]{\LARGE $d\sigma/d p_T\; [\pb/\gev]$}
\psfrag{YJET}{\LARGE $\eta_3$}
\psfrag{PTJET}[c]{\LARGE $p_T\; [\gev]$}
\psfrag{all}{\GNUPlotG{\LARGE all}}
\psfrag{gg}[r]{\GNUPlotA{\LARGE $gg$}}
\psfrag{qg}{\GNUPlotC{\LARGE $qg$}}
\psfrag{qq}{\GNUPlotD{\LARGE $q\bar q$}}

{\setlength{\unitlength}{1cm}
\begin{picture}(7,2.8)(-0.53,.7)
\resizebox{0.6\width}{0.6\height}{
\includegraphics*{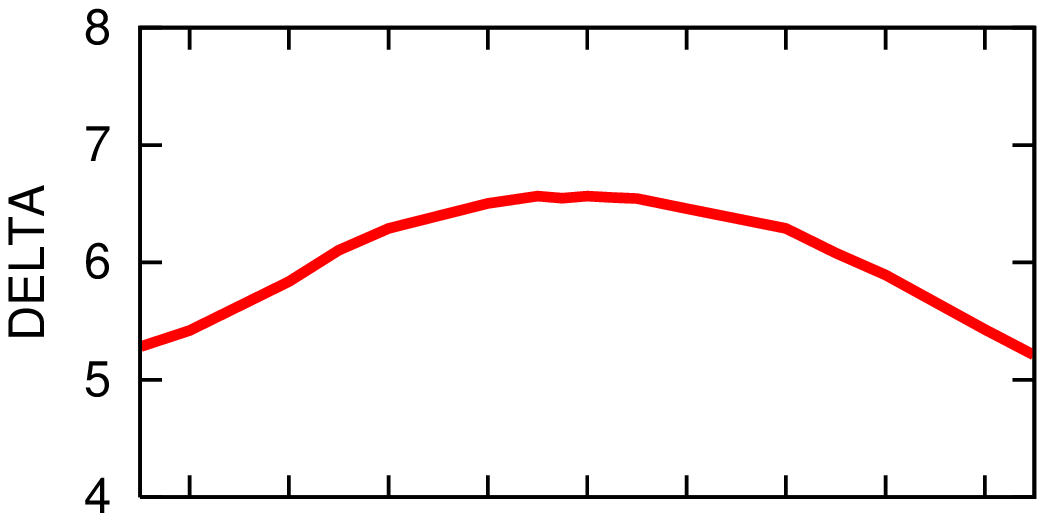}}
\end{picture}
\begin{picture}(7,2.8)(1.53,.7)
\resizebox{0.6\width}{0.6\height}{
\includegraphics*{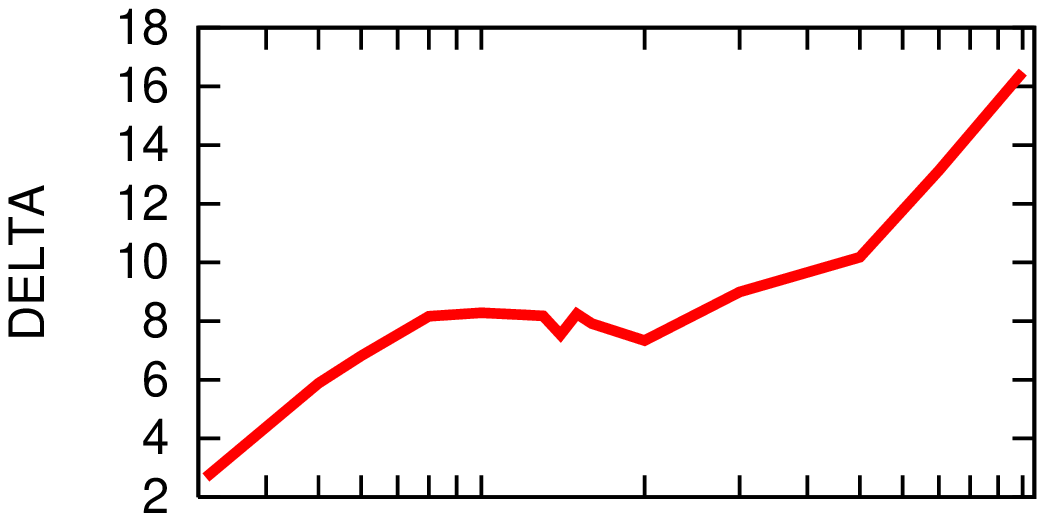}}
\end{picture}
\begin{picture}(7,6)(0,0)
\resizebox{0.6\width}{0.6\height}{
\includegraphics*{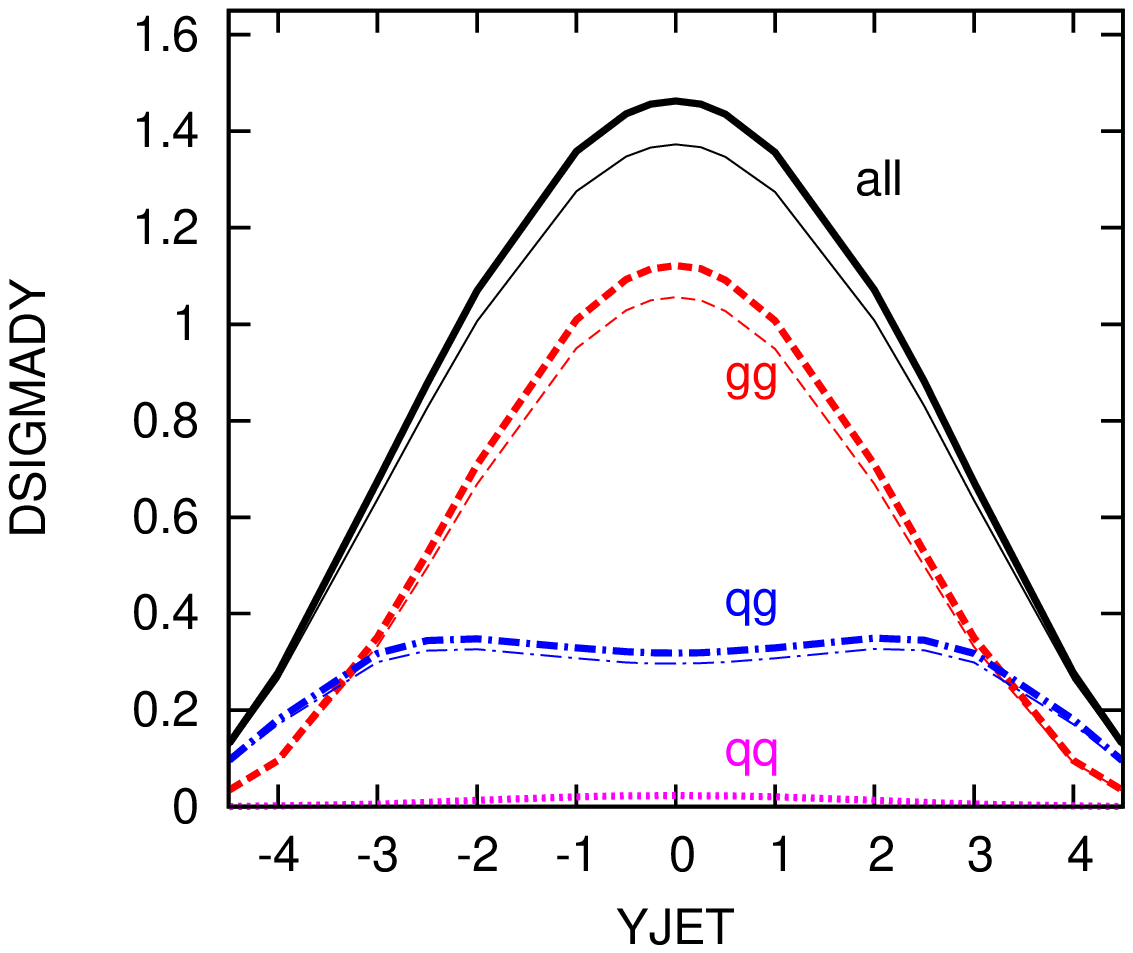}}
\end{picture}
\begin{picture}(7,6)(0,0)
\resizebox{0.6\width}{0.6\height}{
\includegraphics*{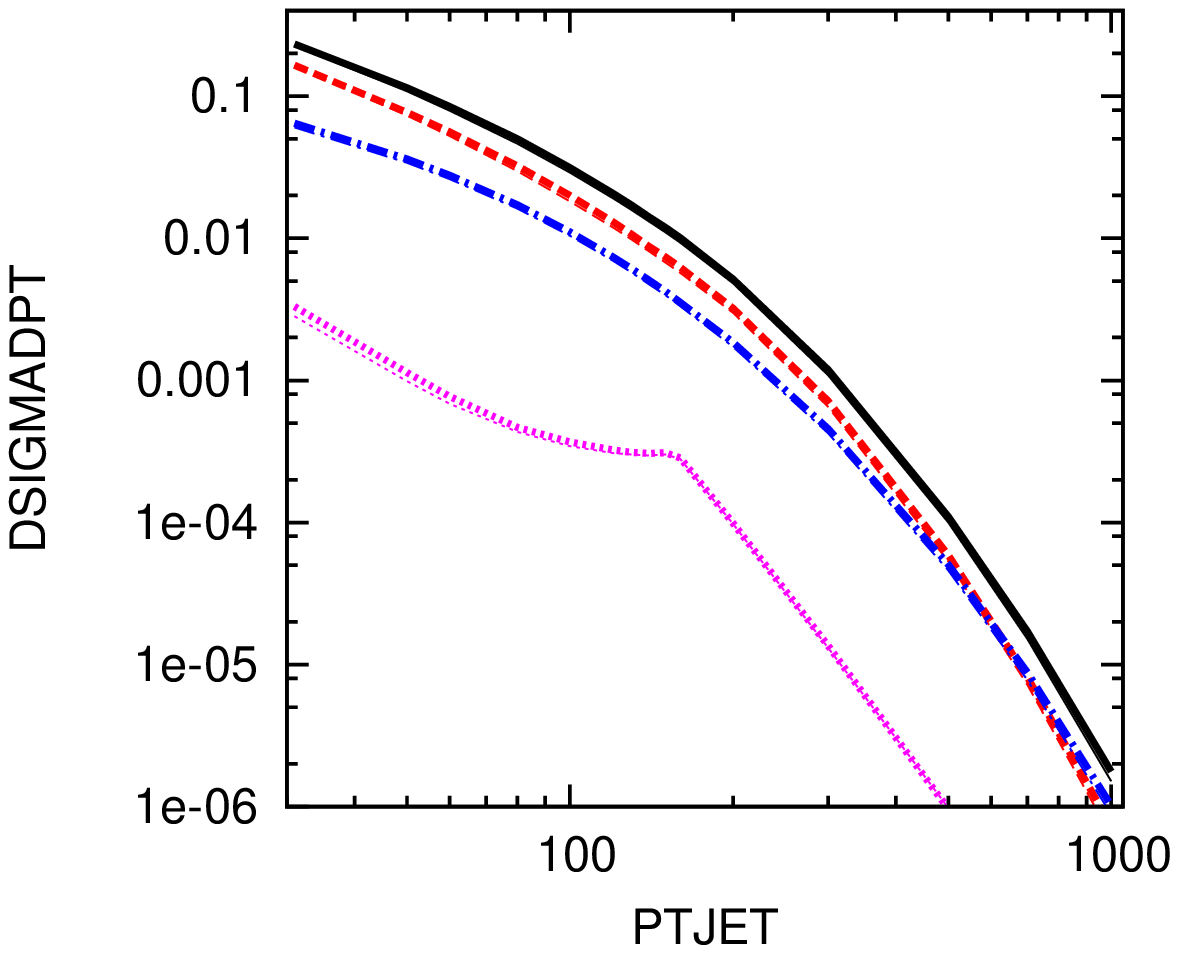}}
\end{picture}
}
\caption{\label{no-mixing} 
LHC, no-mixing(700) scenario with $\ma=500\,\gev$, $\tb=35$:
hadronic cross section for Higgs + jet production 
differential in the jet's pseudorapidity $\eta_3$
and transverse momentum $p_T$ (lower left and right panel).
Thick and thin lines indicate the MSSM and SM predictions
respectively. 
Solid lines indicate the full result, dashed, dot-dashed
and dotted lines the $gg$-, $qg$- and $q \bar q$ contribution
respectively.
In the upper panels the relative difference between the MSSM
and SM result is displayed.}
\end{figure}

\noindent{\bf No-mixing(700) scenario:}\\
Figure~\ref{no-mixing} displays the results for the no-mixing(700) scenario.
The lower left panel shows $d\sigma/d\eta_3$ for the MSSM and SM 
process and also for the three types of subprocesses contributing 
to $h^0$+jet production individually.
The typical two-peak shape of the quark-gluon scattering contribution
is caused by the harder momentum distribution of 
up- and down-type quarks with respect to their anti-particles, 
which lead to a net boost in the
direction of motion of the proton providing the quark,
still visible
in the sum over all partons.
The gluon-fusion and the small $q\bar q$ contribution 
are peaked around $\eta_3=0$.

The upper left panel of Figure~\ref{no-mixing} shows the 
relative difference, $\delta$, between the MSSM and SM prediction for 
$d\sigma/d\eta_3$. 
While the total hadronic cross section in the MSSM is enhanced by
about 6\% compared to the SM, the enhancement of 
$d\sigma/d\eta_3$ varies only by about 1.3\% (between 5.2\% and 6.5\%)
in the range $|\eta_3| < 4.5$.

The right panels in Figure~\ref{no-mixing} show $d\sigma/d p_T$ for the MSSM and SM
processes and the corresponding relative difference, $\delta$.
The thickness of the lines in the lower right panel 
hides the few-percent deviations between MSSM and SM 
for the important contributions from gluon-fusion and quark-gluon scattering.
Interestingly, the deviation between MSSM and SM is 
largest for $q\bar q$ scattering, e.g. 14\% for $p_T=50\,\gev$.
The relative difference in $d\sigma/d p_T$ between the MSSM and the SM
(Figure~\ref{no-mixing}, upper right panel) varies between 3\% for $p_T=50\,\gev$ 
and 8\% (17\%) for $p_T=500\,\gev\;(1\,\tev)$

\begin{figure}[ht]
{\setlength{\unitlength}{1cm}
\begin{picture}(15,4)(1.8,3.5)
\begin{picture}(5.,4)
\put(0,0){\resizebox{0.5\width}{0.5\height}{
\includegraphics*{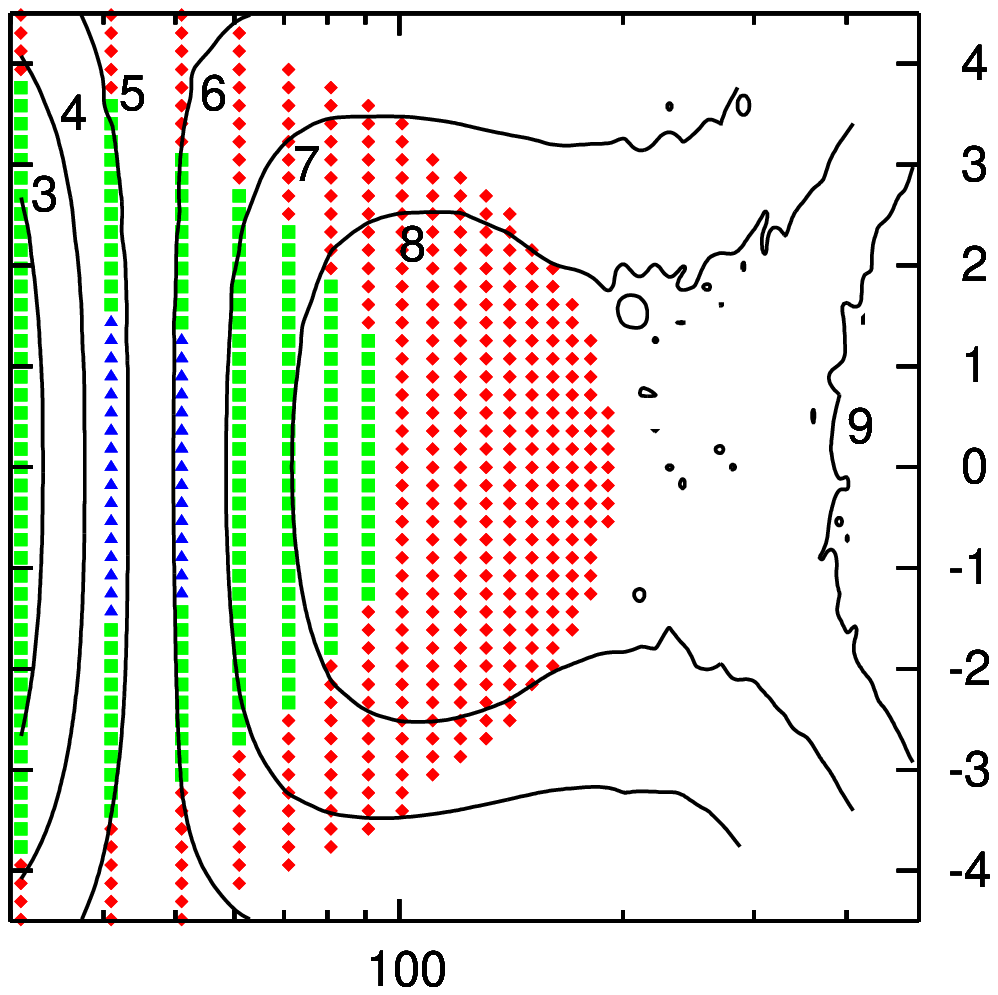}}}
\put(2.3,9.1){\small (a) no-mixing(700)} 
\put(2.3,3.6){$p_T\; [\gev]$}
\put(6.6,9){$\eta_3$}
\end{picture}
\begin{picture}(5.,4)
\put(0,0){\resizebox{0.5\width}{0.5\height}{
\includegraphics*{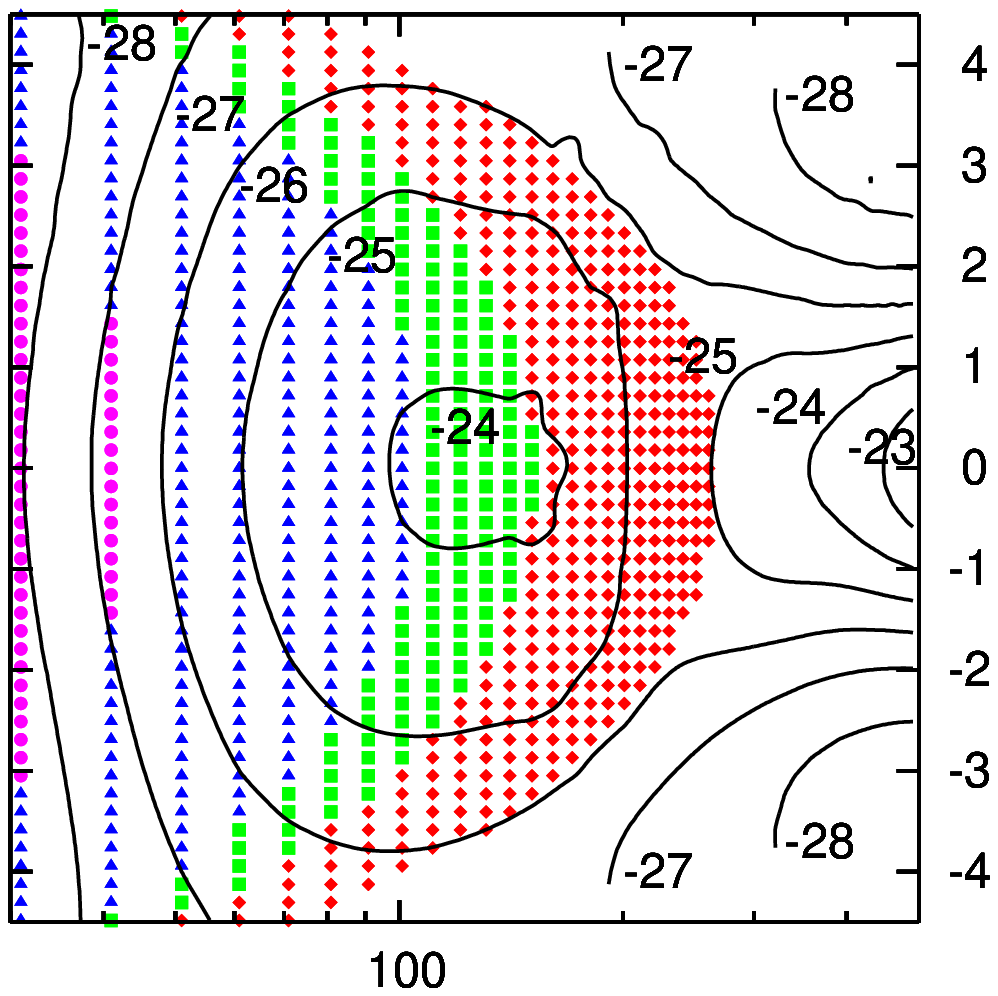}}}
\put(2.3,9.1){\small (b) maximal $m_h$} 
\put(2.3,3.6){$p_T\; [\gev]$}
\put(6.6,9){$\eta_3$}
\end{picture}
\begin{picture}(5.,4)
\put(0,0){\resizebox{0.5\width}{0.5\height}{
\includegraphics*{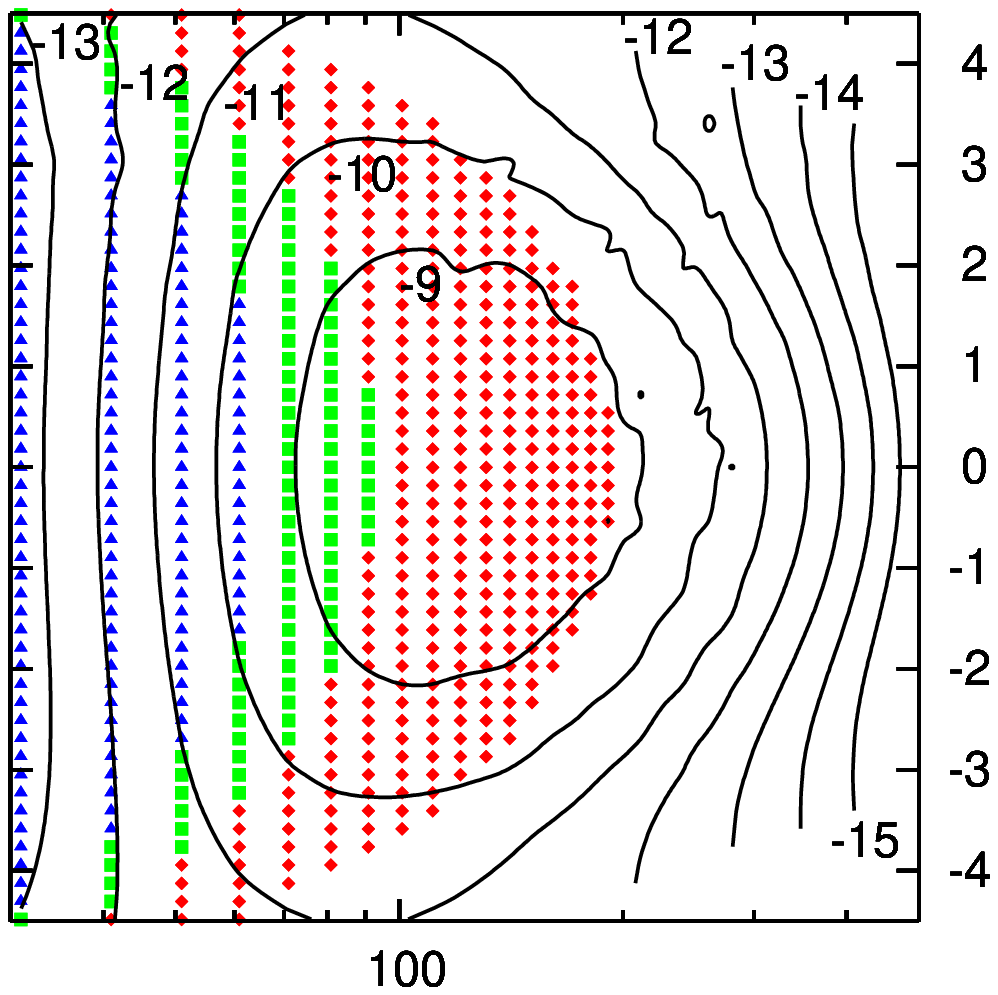}}}
\put(2.3,3.6){$p_T\; [\gev]$}
\put(6.6,9){$\eta_3$}
\put(2.3,9.1){\small (c) small $\alpha_\EFF$} 
\end{picture}
\end{picture}
}
\caption{\label{doubly-diff} Relative and absolute difference between
the MSSM and SM prediction for $d^2\sigma/d\eta_3 d p_T$ at the LHC for the 
three benchmark scenarios as a function of $p_T$ and $\eta_3$.
The contour lines show the relative difference of the predictions in percent, 
while 
diamonds ($\GNUPlotA{\scriptstyle\blacklozenge}$),
squares ($\GNUPlotB{\scriptstyle\blacksquare}$),
triangles ($\GNUPlotC{\scriptstyle\blacktriangle}$),
circles ($\GNUPlotD{\bullet}$),
refer to an absolute difference
in the range 
\GNUPlotA{0.1-0.5 $\fb/\gev$},
\GNUPlotB{0.5-1 $\fb/\gev$},
\GNUPlotC{1-5 $\fb/\gev$},
\GNUPlotD{5-10 $\fb/\gev$}
respectively.
In the white area the difference is less than $0.1\,\fb/\gev$.
}
\end{figure}

The relative difference between the MSSM and SM prediction
for the two-fold differential cross section $d^2\sigma/d p_T/d\eta_3$,
indicated by the contours in Figure~\ref{doubly-diff}(a), 
shows a non-trivial behaviour
with an overall variation of more than 6\% in the depicted range,
$|\eta_3| < 4.5$ and $30\,\gev < p_T <500\,\gev$.
The differently shaped dots in Figure~\ref{doubly-diff}(a) 
show the absolute difference between MSSM and SM, 
which gives an idea of the kinematical region where the LHC 
experiments may become sensitive to this difference.

Modifying the cuts may increase the sensitivity 
to deviations from the SM.
Guided by Figure~\ref{doubly-diff}, we study 
the cross section $\sigma_f$ with rather soft forwardish jets
and $\sigma_c$ with harder more central jets:
\begin{align}
\label{sigmac}
\sigma_c & = \sigma\left(pp\to h^0 + \JET +X \right)|_{|\eta| < 1.5,\; p_T > 70\,\gev}
\,,\\
\sigma_f & = \sigma\left(pp\to h^0 + \JET +X \right)|_{1.5 < |\eta| < 4.5,\; 30\,\gev < p_T < 50\,\gev}
\,.
\end{align}
The results are put together in Table~\ref{thetable}, where also the ratio
\begin{align}
\label{ratio}
R & = \frac{\sigma_c}{\sigma_f}\,.
\end{align}
and the relative difference between MSSM and SM
\begin{align}
\Delta & = \frac{R_\MSSM- R_\SM}{R_\SM}
\end{align}
are listed.
While each individual cross section in the MSSM and the SM
is still of the order of $1\,\pb$, 
which translates into $10^5$ raw events for an integrated luminosity 
of $100\,\fb^{-1}$, 
the MSSM ratio $R_\MSSM$ differs by $4.5\%$ compared to $R_\SM$.

\begin{table}[t]
\begin{center}
\begin{tabular}{c|cc|cc|cc}
quantity & \multicolumn{2}{c|}{no-mixing(700)} & \multicolumn{2}{c|}{$m_h^\MAX(400)$} 
        & \multicolumn{2}{c}{small $\alpha_{\text{eff}}$} \\
         & SM & MSSM & SM & MSSM & SM & MSSM\\
\hline
$\sigma_c$
 & 1.623$\,\pb$ & 1.762$\,\pb$ & 1.448$\,\pb$ & 1.096$\,\pb$ & 1.490$\,\pb$ & 1.356$\,\pb$ \\
$\sigma_f$
 & 1.682$\,\pb$ & 1.749$\,\pb$ & 1.419$\,\pb$ & 1.031$\,\pb$ & 1.480$\,\pb$ & 1.299$\,\pb$\\
\hline
$R = \sigma_c/\sigma_f$ 
 & 0.965 & 1.008 & 1.020 & 1.063 & 1.007 & 1.044\\
\hline
$\Delta$ 
&\multicolumn{2}{c|}{4.5\%}
        &\multicolumn{2}{c|}{4.2\%}&\multicolumn{2}{c}{3.7\%} 
\end{tabular}
\end{center}
\caption{\label{thetable} Cross section predictions 
for Higgs + jet production with jets radiated into the central ($\sigma_c$) 
and forward part of the detector ($\sigma_f$), together with
their ratio $R$ and the relative difference between the MSSM and 
SM value for $R$, $\Delta$.
}
\end{table}
\smallskip

\begin{figure}[ht]
\psfrag{DELTA}{\LARGE $\delta\; [\%]$}
\psfrag{DSIGMADY}[t][b]{\LARGE $d\sigma/d\eta_3\; [\pb]$}
\psfrag{DSIGMADPT}[t][b]{\LARGE $d\sigma/d p_T\; [\pb/\gev]$}
\psfrag{YJET}{\LARGE $\eta_3$}
\psfrag{PTJET}[c]{\LARGE $p_T\; [\gev]$}
\psfrag{all}{\GNUPlotG{\LARGE all}}
\psfrag{gg}[c]{\GNUPlotA{\LARGE $gg$}}
\psfrag{qg}[c]{\GNUPlotC{\LARGE $qg$}}
\psfrag{qq}{\GNUPlotD{\LARGE $q\bar q$}}

{\setlength{\unitlength}{1cm}
\begin{picture}(7,2.8)(-.18,.7)
\resizebox{0.6\width}{0.6\height}{
\includegraphics*{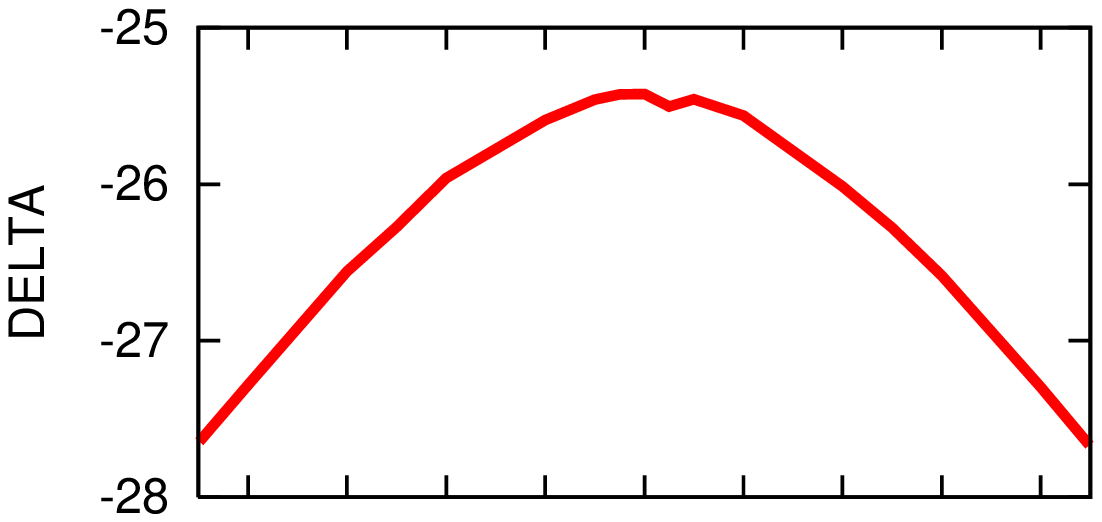}}
\end{picture}
\begin{picture}(7,2.8)(1.53,.7)
\resizebox{0.6\width}{0.6\height}{
\includegraphics*{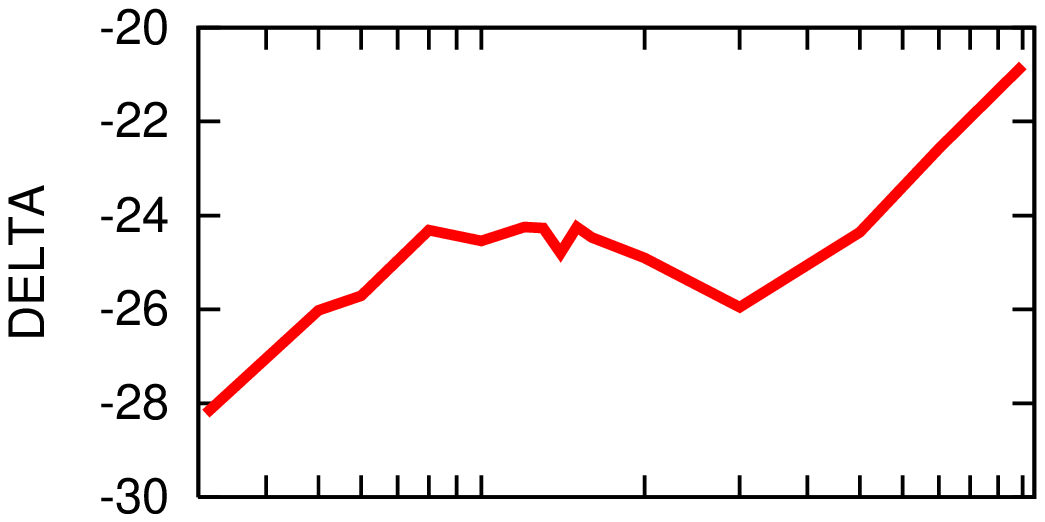}}
\end{picture}
\begin{picture}(7,6)(0,0)
\resizebox{0.6\width}{0.6\height}{
\includegraphics*{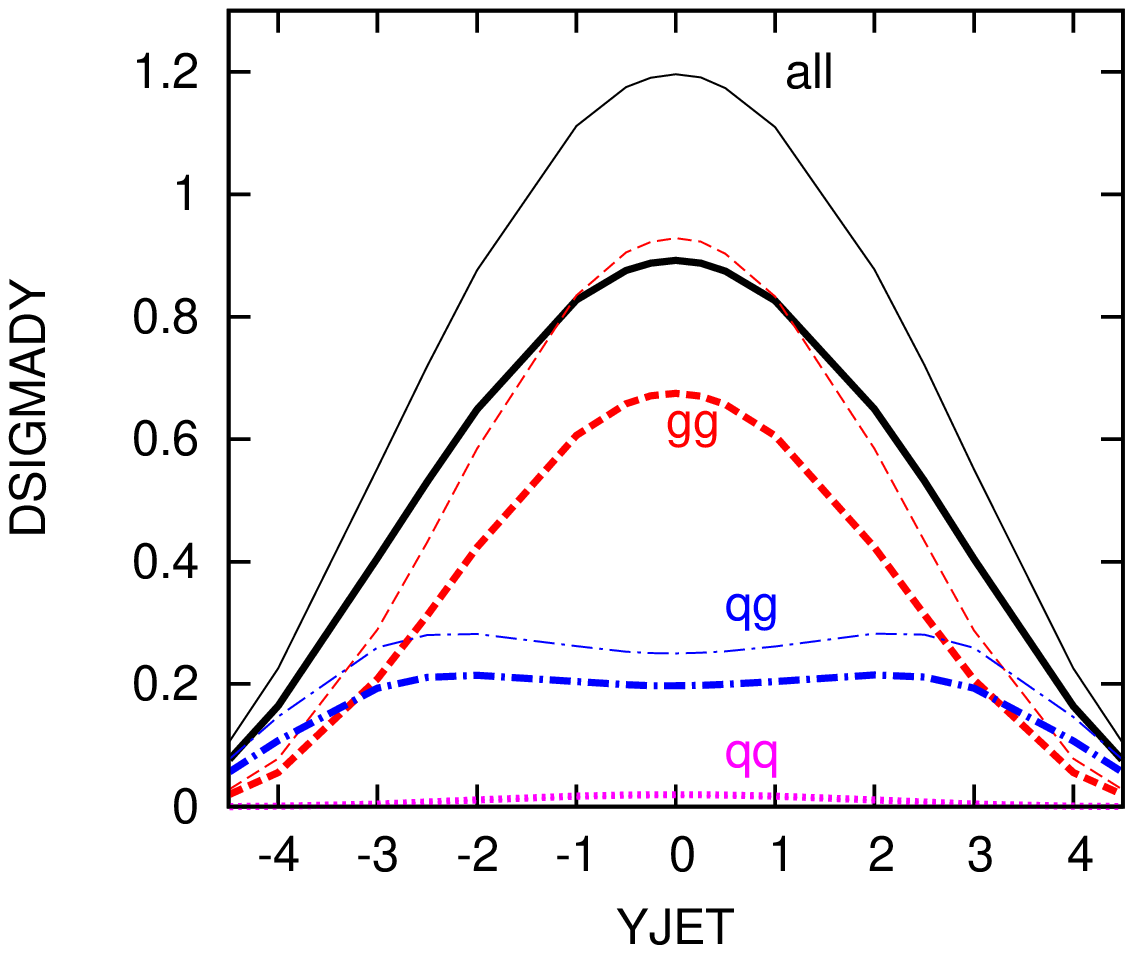}}
\end{picture}
\begin{picture}(7,6)(0,0)
\resizebox{0.6\width}{0.6\height}{
\includegraphics*{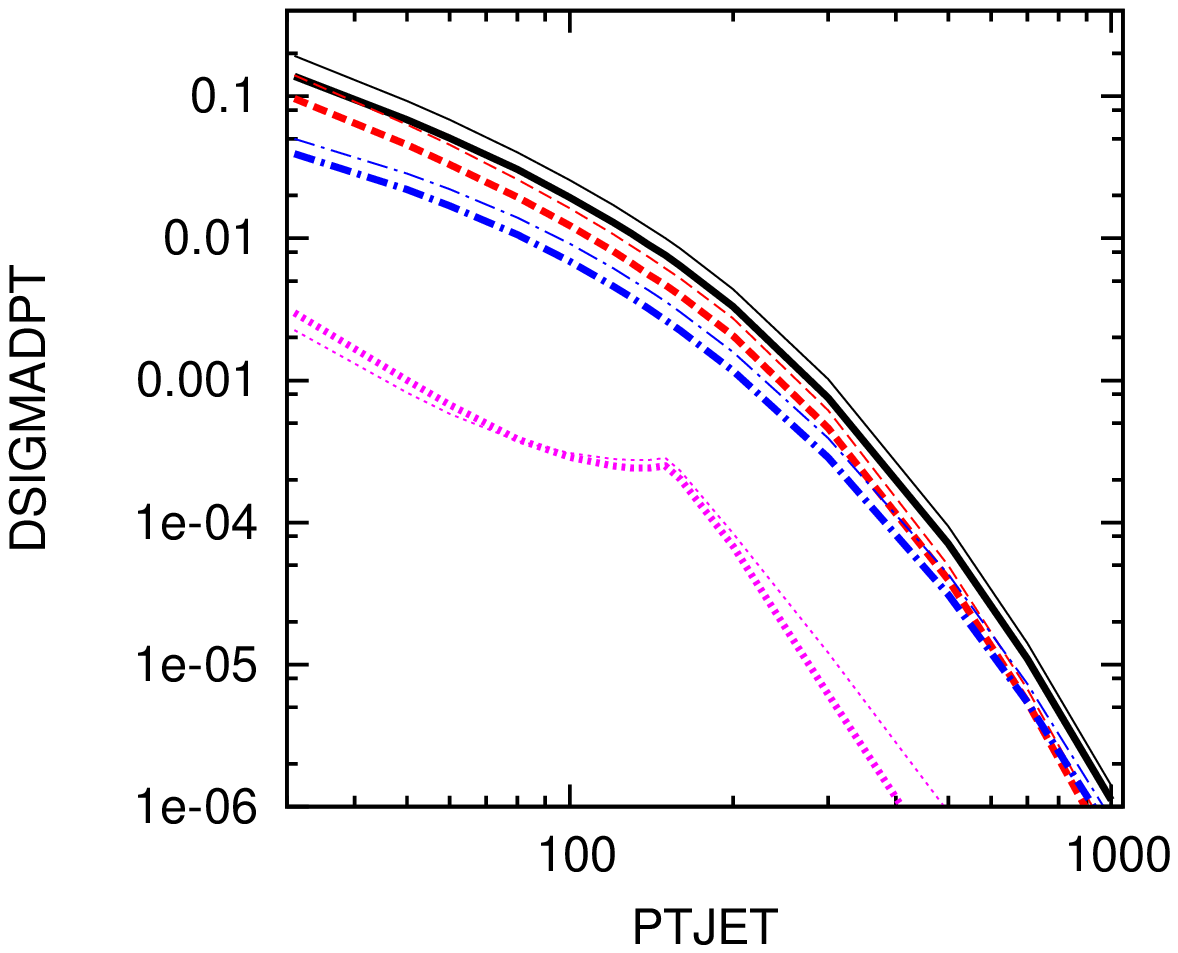}}
\end{picture}
}
\caption{\label{mh-max} LHC, mh-max scenario with $\ma=400\,\gev$, $\tb=30$ :
differential hadronic cross sections for Higgs + jet production .
See caption of Figure~\ref{no-mixing} for more details.
}
\end{figure}

\noindent{\bf {\boldmath{$m_h^\MAX(400)$}} scenario:}\\
Figure \ref{mh-max} contains
$d\sigma/d\eta_3$ and $d\sigma/d p_T$
for the $m_h^\MAX(400)$ scenario and the corresponding relative differences to
the SM prediction.
The total hadronic cross section in this MSSM scenario 
is about 25\% smaller than in the SM. 
Yet, as far as the $\eta_3$ and $p_T$ dependent differences between
MSSM and SM are concerned, the same qualitative picture appears.
The variation of the relative difference $\delta$ with $\eta_3$ 
in the range $|\eta_3| < 4.5$ is about 2\% and 
with $p_T$ in the range $p_T \in [30\,\gev,1000\,\gev]$
is about 7\%.

Similar to the no-mixing(700) scenario, 
the difference in the doubly differential cross section
in Figure~\ref{doubly-diff}(b) has a non-trivial $\eta_3$ and 
$p_T$ dependence.
This suggests a similar refinement as in 
the no-mixing(700) scenario.
For simplicity, we calculate the same ratio of cross sections according
to the same cuts as before (see Eqs.~(\ref{sigmac}) to (\ref{ratio}))
for the SM and MSSM.
Table~\ref{thetable} shows $\Delta = 4.2\%$ as
the relative difference of the ratios. 

\smallskip

\begin{figure}[ht]
\psfrag{DELTA}{\LARGE $\delta\; [\%]$}
\psfrag{DSIGMADY}[t][b]{\LARGE $d\sigma/d\eta_3\; [\pb]$}
\psfrag{DSIGMADPT}[t][b]{\LARGE $d\sigma/d p_T\; [\pb/\gev]$}
\psfrag{YJET}{\LARGE $\eta_3$}
\psfrag{PTJET}[c]{\LARGE $p_T\; [\gev]$}
\psfrag{all}{\GNUPlotG{\LARGE all}}
\psfrag{gg}[r][t]{\GNUPlotA{\LARGE $gg$}}
\psfrag{qg}[c]{\GNUPlotC{\LARGE $qg$}}
\psfrag{qq}{\GNUPlotD{\LARGE $q\bar q$}}

{\setlength{\unitlength}{1cm}
\begin{picture}(7,2.8)(-.18,.7)
\resizebox{0.6\width}{0.6\height}{
\includegraphics*{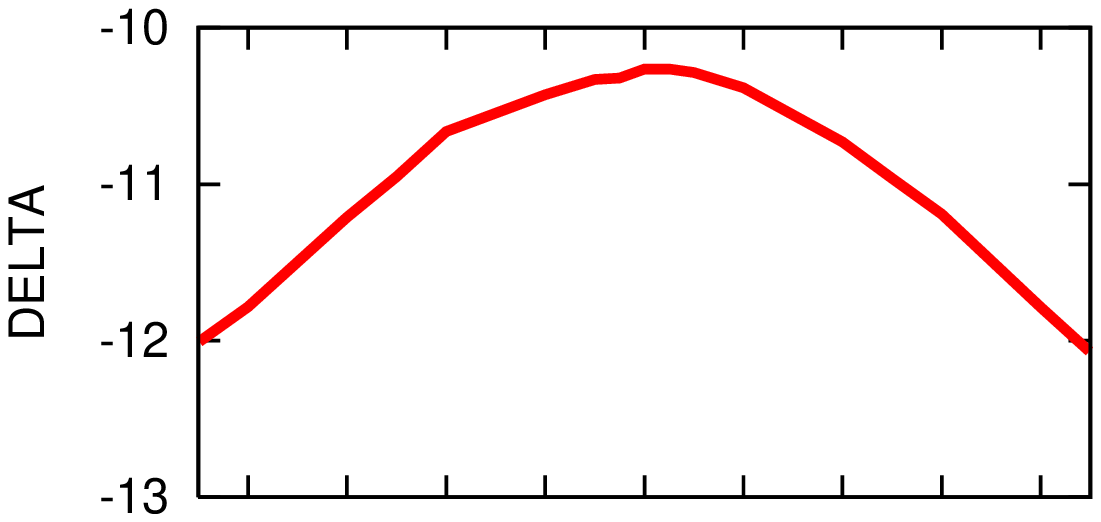}}
\end{picture}
\begin{picture}(7,2.8)(1.53,.7)
\resizebox{0.6\width}{0.6\height}{
\includegraphics*{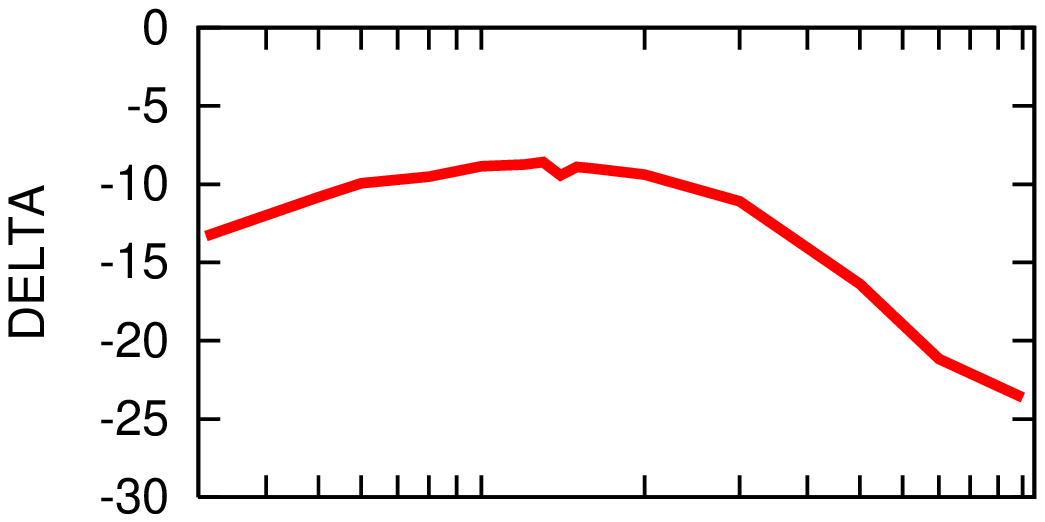}}
\end{picture}
\begin{picture}(7,6)(0,0)
\resizebox{0.6\width}{0.6\height}{
\includegraphics*{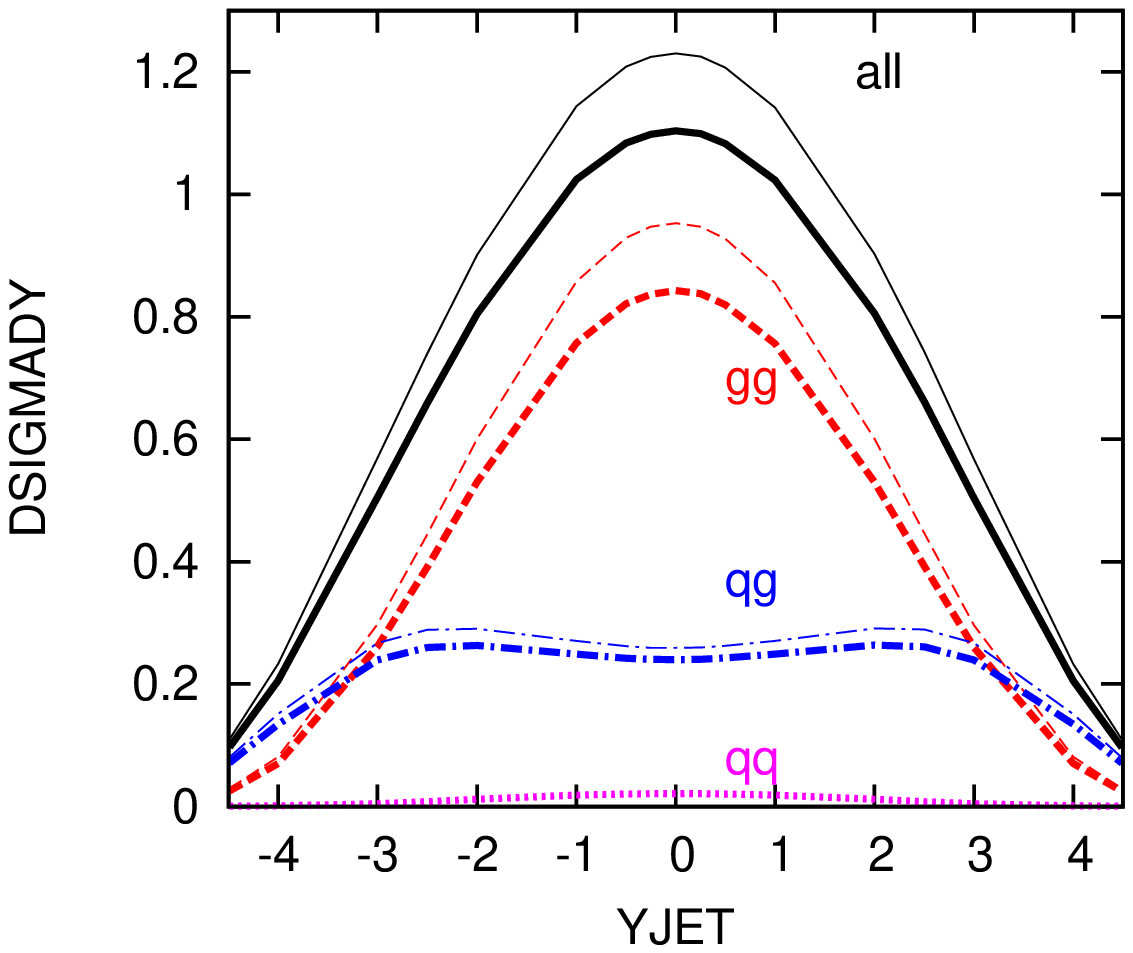}}
\end{picture}
\begin{picture}(7,6)(0,0)
\resizebox{0.6\width}{0.6\height}{
\includegraphics*{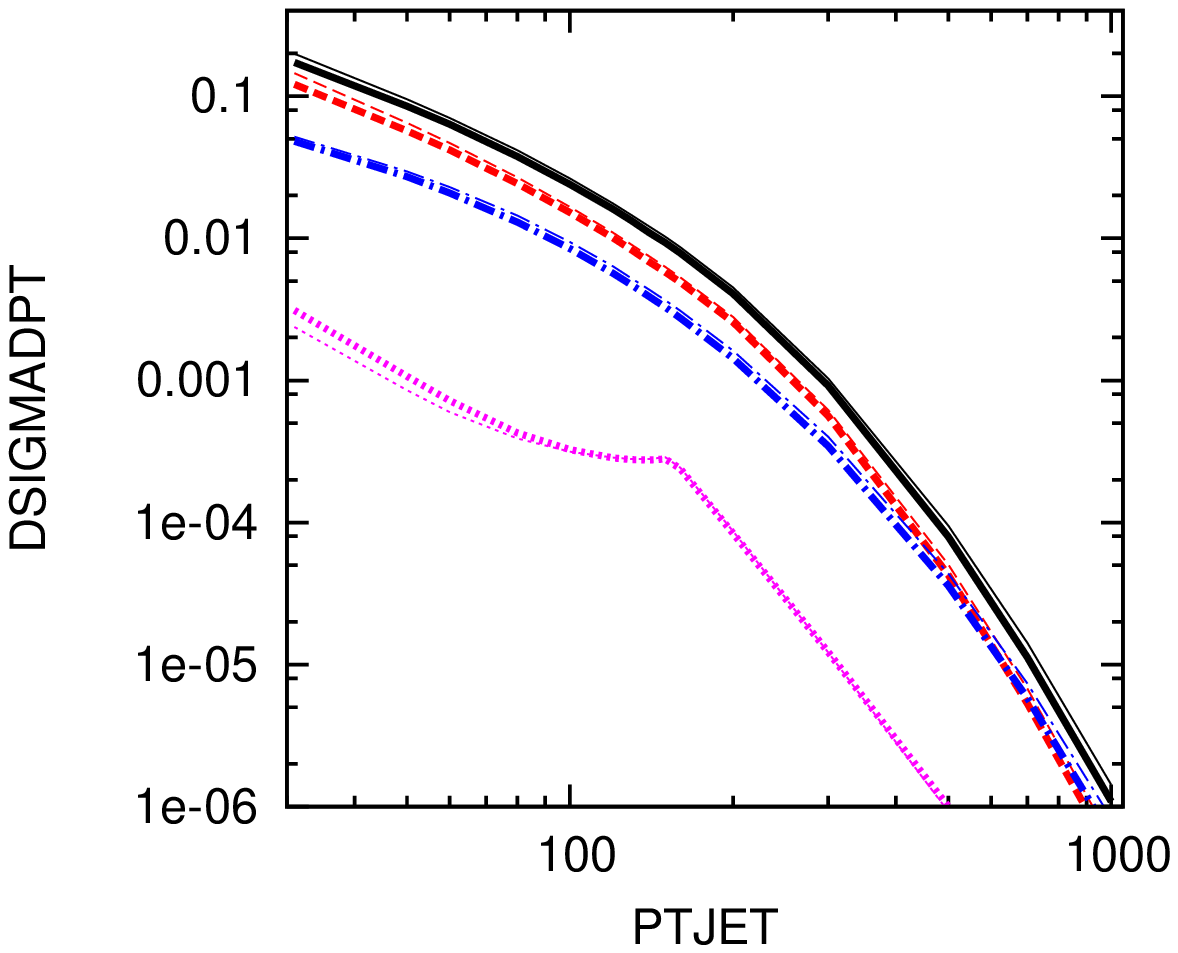}}
\end{picture}
}
\caption{\label{smalla} LHC, small alpha-eff scenario with $\ma=400\,\gev$, $\tb=30$:
differential hadronic cross sections for Higgs + jet production.
See caption of Figure~\ref{no-mixing} for more details.
}
\end{figure}

\noindent{\bf Small-{\boldmath{$\alpha_{\text{eff}}$}} scenario:}\\
In Figure \ref{smalla}, 
$d\sigma/d\eta_3$ and $d\sigma/d p_T$ for the 
small-$\alpha_{\text{eff}}$ scenario 
and corresponding relative differences to the SM
are shown.
While the total hadronic cross section in this scenario
is about 11\% below the SM, the variation in the $\eta_3$ distribution
is about 2\%, as in the other two scenarios.
Opposite to the other two scenarios the $p_T$ spectrum is
slightly softer than in the SM.
The range of variation of $d\sigma/d p_T$ with $p_T$
is about $15\%$.

The doubly differential cross section
in Figure~\ref{doubly-diff}(c) shows a behaviour similar to the
other two scenarios. Calculating the cross section ratio $R$
according to Eqs.~(\ref{sigmac}) to (\ref{ratio}) 
we get $\Delta = 3.7\%$ (see Table~\ref{thetable}).

\subsubsection{Low {\boldmath{$m_A$}}}

\begin{figure}[ht]
\psfrag{DELTA}{\LARGE $\delta\; [1]$}
\psfrag{DSIGMADY}[t][b]{\LARGE $d\sigma/d\eta_3\; [\pb]$}
\psfrag{DSIGMADPT}[t][b]{\LARGE $d\sigma/d p_T\; [\pb/\gev]$}
\psfrag{YJET}{\LARGE $\eta_3$}
\psfrag{PTJET}[c]{\LARGE $p_T\; [\gev]$}
\psfrag{all}{\GNUPlotG{\LARGE all}}
\psfrag{gg}[c]{\GNUPlotA{\LARGE $gg$}}
\psfrag{qg}{\GNUPlotC{\LARGE $qg$}}
\psfrag{qq}[b]{\GNUPlotD{\LARGE $q\bar q$}}

{\setlength{\unitlength}{1cm}
\begin{picture}(7,2.8)(0,.7)
\resizebox{0.6\width}{0.6\height}{
\includegraphics*{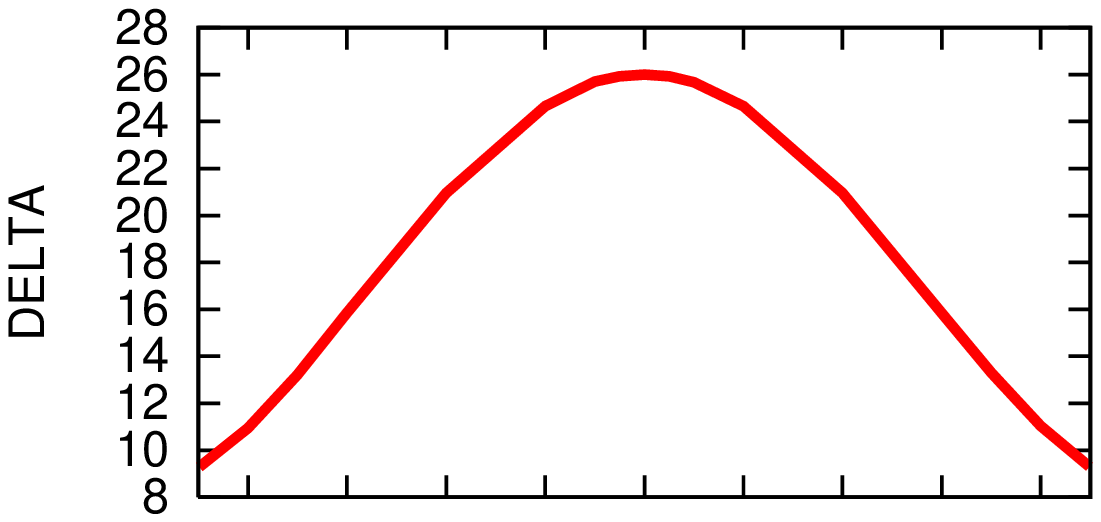}}
\end{picture}
\begin{picture}(7,2.8)(1.53,.7)
\resizebox{0.6\width}{0.6\height}{
\includegraphics*{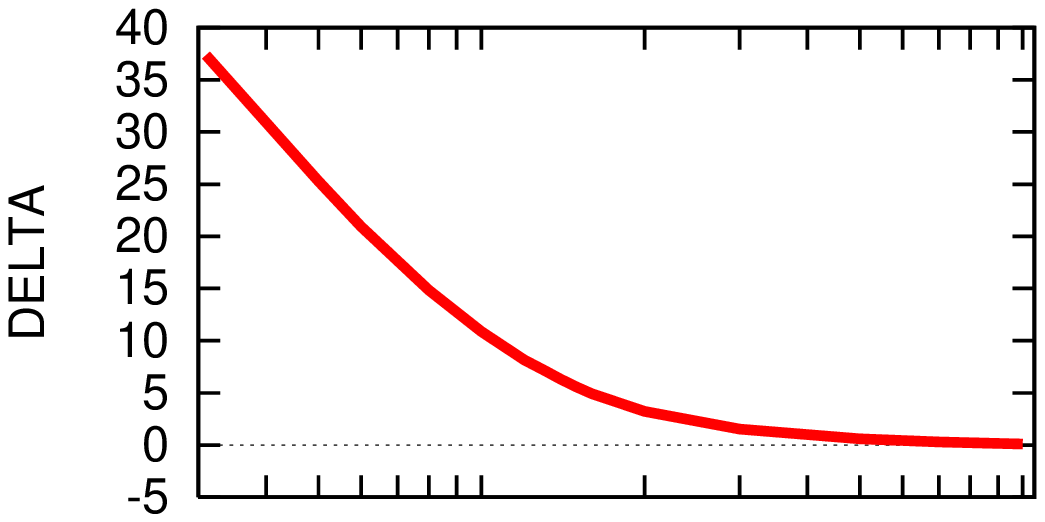}}
\end{picture}
\begin{picture}(7,6)(0,0)
\resizebox{0.6\width}{0.6\height}{
\includegraphics*{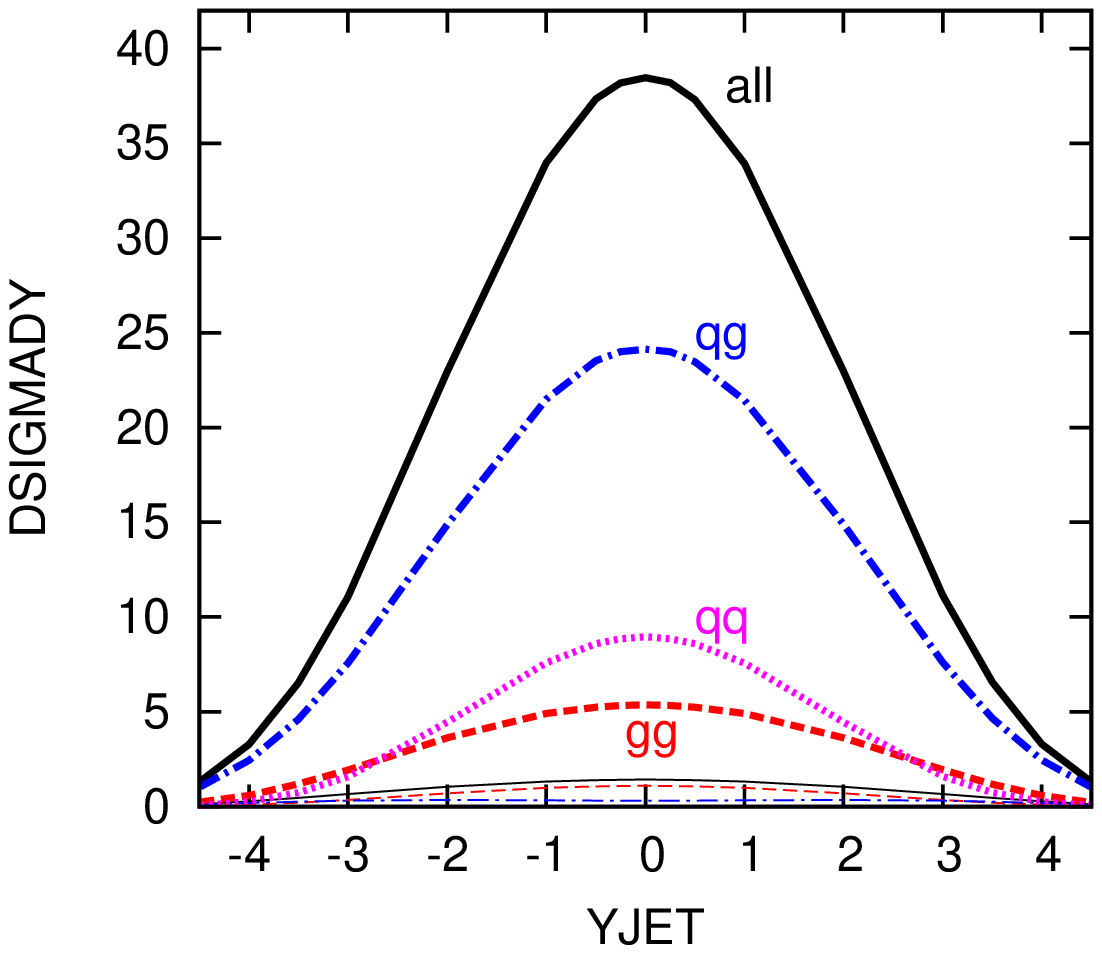}}
\end{picture}
\begin{picture}(7,6)(0,0)
\resizebox{0.6\width}{0.6\height}{
\includegraphics*{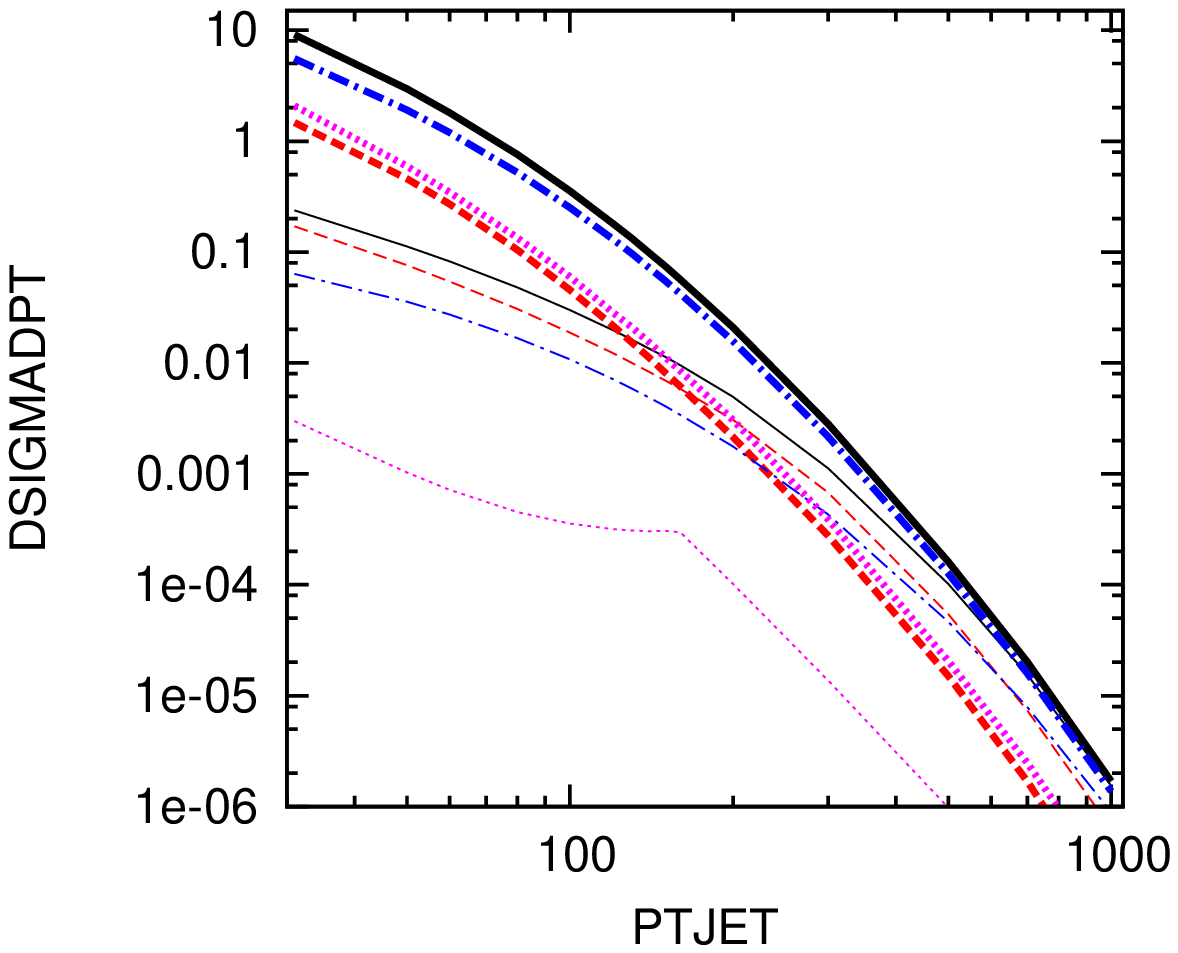}}
\end{picture}
}
\caption{\label{mh-max-smallma} LHC, mh-max scenario with $\ma=110\,\gev$, $\tb=30$:
differential hadronic cross sections for Higgs + jet production.
See caption of Figure~\ref{no-mixing} for more details.}
\end{figure}

As an example for the low-$m_A$ case at the LHC we show results for 
the $m_h^\MAX(400)$ scenario in Figure~\ref{mh-max-smallma}.
The change with respect to the SM is dramatic.
Due to the enhanced cross sections of the $b$-quark processes,
the quark-gluon scattering contribution dominates the cross section
and even the contribution from $q \bar q$ is larger than from gluon fusion.
The total hadronic cross section in the MSSM is 22 times higher 
than in the SM ($\approx 175\,\pb$).

Out of all jets allowed by our cuts~(\ref{the-cuts})
a larger fraction
of jets is radiated into the central part of the detector.
For instance, the fraction of jets 
produced with $|\eta_3| < 2$ compared 
to all jets allowed by the cuts 
is 93\% in the MSSM versus 85\% in the SM.
Correspondingly, the $p_T$ spectrum is much softer than in the SM,
yielding an enhanced rate for processes with jet transverse momenta 
below $850\,\gev$, e.g. by a factor of 10
for $p_T=100\,\gev$, and rates similar to the SM above $850\,\gev$.

\subsection{Differential cross sections at the Tevatron}

The typical hadronic cross section for Higgs + jet in the SM
expected at the Tevatron for the cuts $p_T > 30\,\gev$ and $|\eta_3| < 4.5$
is around $0.1\,\pb$ for Higgs masses around 100\,\gev, 
which is possibly not sufficient to be observable at the Tevatron.
Therefore, for the Tevatron only the MSSM scenarios with low
$m_A$ and $\tb$ not too small are of interest.
Those scenarios exhibit a cross section  enhanced by a factor of 
up to 30 compared to the SM \cite{hjet-own}.
This is due to the contribution of $b$-quark initiated processes
which become dominant because of the strongly enhanced Yukawa coupling 
of $b$-quarks to the Higgs boson $h^0$.

\begin{figure}[ht]
\psfrag{DELTA}{\LARGE $\delta\; [1]$}
\psfrag{DSIGMADY}[t][b]{\LARGE $d\sigma/d\eta_3\; [\pb]$}
\psfrag{DSIGMADPT}[t][b]{\LARGE $d\sigma/d p_T\; [\pb/\gev]$}
\psfrag{YJET}{\LARGE $\eta_3$}
\psfrag{PTJET}[c]{\LARGE $p_T\; [\gev]$}
\psfrag{all}{\GNUPlotG{\LARGE all}}
\psfrag{gg}[c]{\GNUPlotA{\LARGE $gg$}}
\psfrag{qg}{\GNUPlotC{\LARGE $qg$}}
\psfrag{qq}[b]{\GNUPlotD{\LARGE $q\bar q$}}

{\setlength{\unitlength}{1cm}
\begin{picture}(7,2.8)(-.18,.7)
\resizebox{0.6\width}{0.6\height}{
\includegraphics*{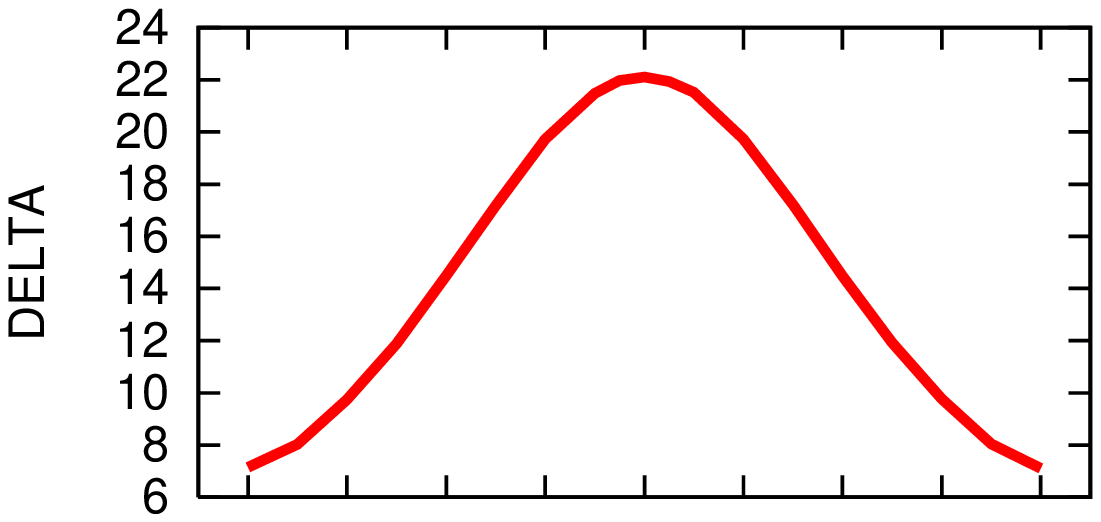}}
\end{picture}
\begin{picture}(7,2.8)(1.53,.7)
\resizebox{0.6\width}{0.6\height}{
\includegraphics*{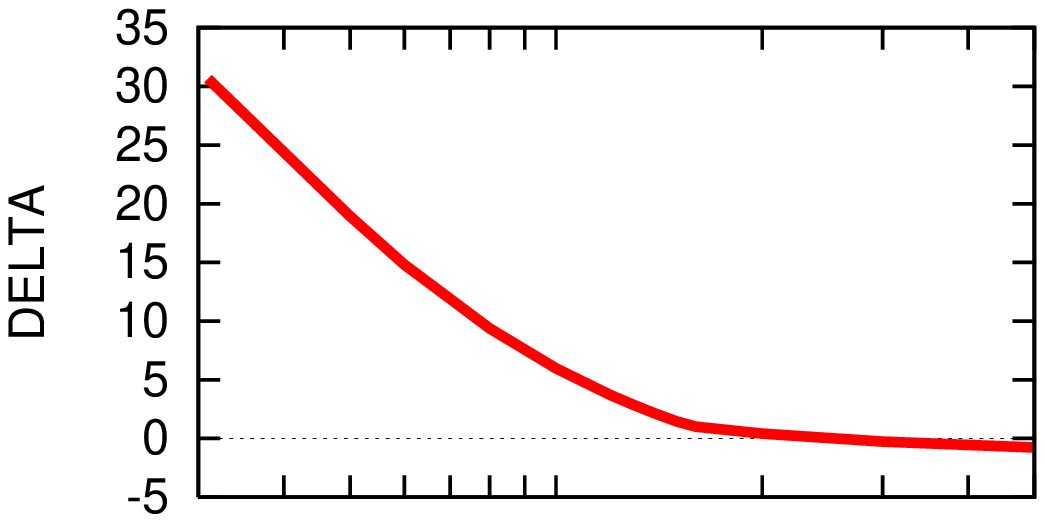}}
\end{picture}
\begin{picture}(7,6)(0,0)
\resizebox{0.6\width}{0.6\height}{
\includegraphics*{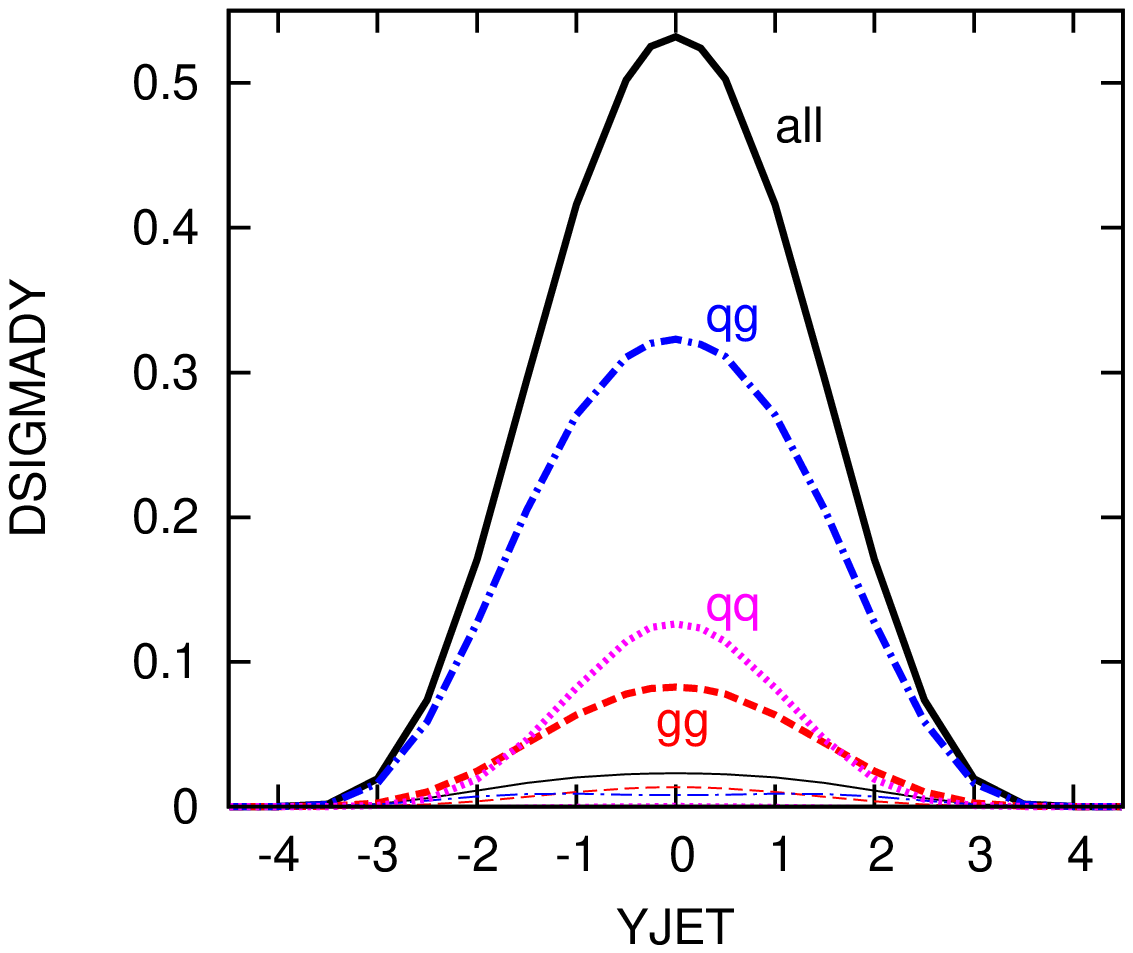}}
\end{picture}
\begin{picture}(7,6)(0,0)
\resizebox{0.6\width}{0.6\height}{
\includegraphics*{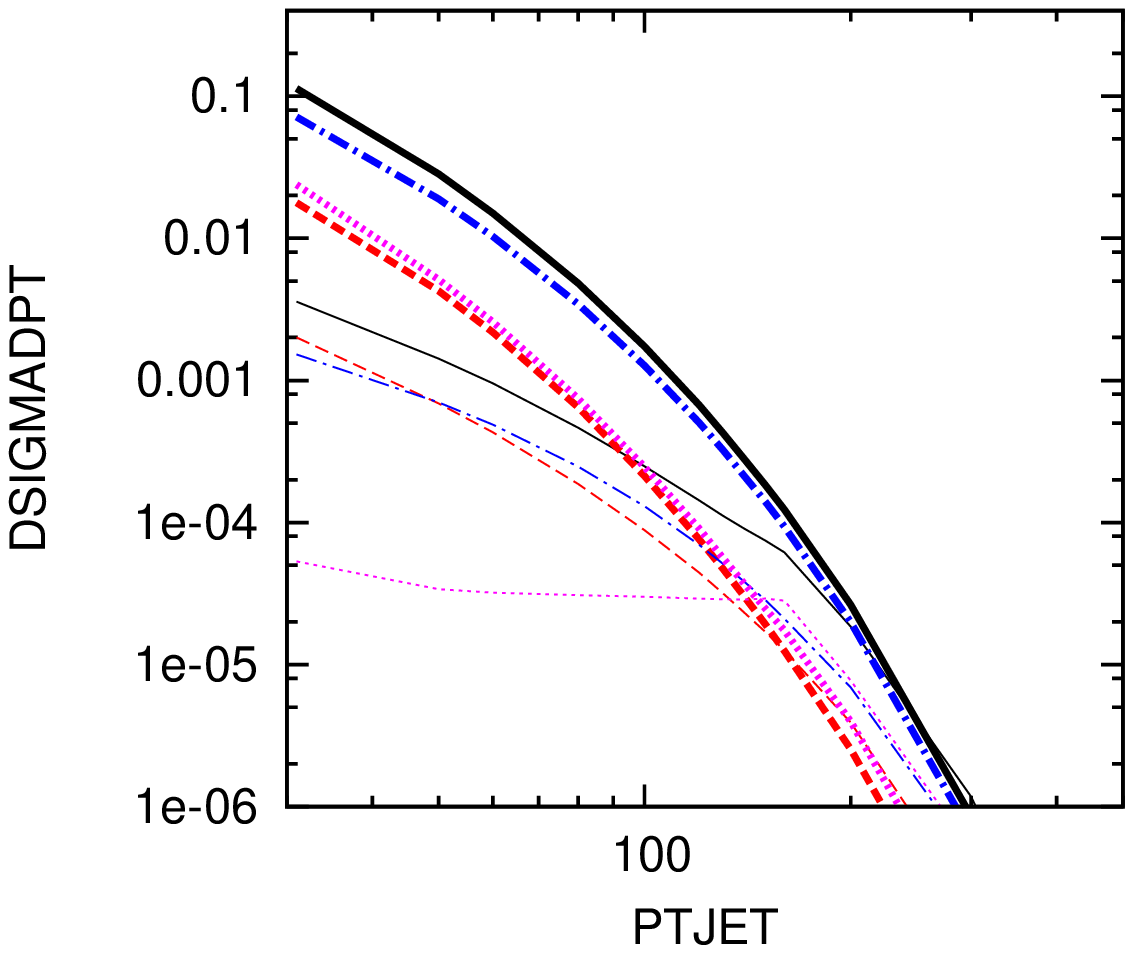}}
\end{picture}
}
\caption{\label{mh-max-smallma-tev} Tevatron, mh-max scenario with $\ma=110\,\gev$, $\tb=30$:
differential hadronic cross sections for Higgs + jet production.
See caption of Figure~\ref{no-mixing} for more details.}
\end{figure}

Figure~\ref{mh-max-smallma-tev} shows results for 
the same low-$m_A$ scenario as just described for the LHC in the 
previous paragraph.
Very similar to the LHC case, we see a strongly enhanced total
hadronic cross section with a softer $p_T$ spectrum and a larger 
fraction of jets radiated into the central part of the detector than in the SM.

\section{Conclusions}

We have calculated pseudorapidity and transverse momentum 
distributions
for the MSSM $h^0$ + high-$p_T$ jet production cross section
at the LHC and the Tevatron.
For scenarios with large $m_A$,
the loop-induced processes dominate the cross section, and
superpartners can have a significant impact when they are not too heavy.
For small $m_A$, the Yukawa couplings of the $b$-quarks are enhanced and 
hence the cross section is dominated by $b$-quark induced 
tree-level parton reactions.
The example investigated here,
the $m_h^\MAX(400)$ scenario, 
shows a strongly enhanced hadronic cross section compared to the SM,
by a factor of more than 20.
Such a scenario predicts for both LHC and Tevatron
a softer $p_T$ spectrum, with a fraction
of jets radiated into the central part of the detector
larger than in the SM.

\section*{Acknowledgement} 
This work was supported in part by the European Community's 
Marie-Curie Research
Training Network under contract MRTN-CT-2006-035505
`Tools and Precision Calculations for Physics Discoveries at Colliders'
(HEPTOOLS).

\newpage

\begin{figure}[tb]
\begin{center}
\begin{footnotesize}
\unitlength=1bp%

\begin{feynartspicture}(432,90)(4,1)

\FADiagram{}
\FAProp(0.,15.)(10.,14.5)(0.,){/Straight}{1}
\FALabel(4.9226,13.6819)[t]{$u$}
\FAProp(0.,5.)(6.,5.5)(0.,){/Cycles}{0}
\FALabel(3.12872,4.18535)[t]{$g$}
\FAProp(20.,15.)(10.,14.5)(0.,){/Straight}{-1}
\FALabel(14.9226,15.8181)[b]{$u$}
\FAProp(20.,5.)(14.,5.5)(0.,){/ScalarDash}{0}
\FALabel(16.892,4.43449)[t]{$h^0$}
\FAProp(10.,14.5)(10.,11.5)(0.,){/Cycles}{0}
\FALabel(11.77,13.)[l]{$g$}
\FAProp(6.,5.5)(14.,5.5)(0.,){/Straight}{-1}
\FALabel(10.,4.43)[t]{$q_i$}
\FAProp(6.,5.5)(10.,11.5)(0.,){/Straight}{1}
\FALabel(7.19032,8.87979)[br]{$q_i$}
\FAProp(14.,5.5)(10.,11.5)(0.,){/Straight}{-1}
\FALabel(12.8097,8.87979)[bl]{$q_i$}
\FAVert(10.,14.5){0}
\FAVert(6.,5.5){0}
\FAVert(14.,5.5){0}
\FAVert(10.,11.5){0}

\FADiagram{}
\FAProp(0.,15.)(10.,14.5)(0.,){/Straight}{1}
\FALabel(4.9226,13.6819)[t]{$u$}
\FAProp(0.,5.)(6.,5.5)(0.,){/Cycles}{0}
\FALabel(3.12872,4.18535)[t]{$g$}
\FAProp(20.,15.)(10.,14.5)(0.,){/Straight}{-1}
\FALabel(14.9226,15.8181)[b]{$u$}
\FAProp(20.,5.)(14.,5.5)(0.,){/ScalarDash}{0}
\FALabel(16.892,4.43449)[t]{$h^0$}
\FAProp(10.,14.5)(10.,11.5)(0.,){/Cycles}{0}
\FALabel(11.77,13.)[l]{$g$}
\FAProp(6.,5.5)(14.,5.5)(0.,){/ScalarDash}{-1}
\FALabel(10.,4.43)[t]{$\tilde q_i^s$}
\FAProp(6.,5.5)(10.,11.5)(0.,){/ScalarDash}{1}
\FALabel(7.19032,8.87979)[br]{$\tilde q_i^s$}
\FAProp(14.,5.5)(10.,11.5)(0.,){/ScalarDash}{-1}
\FALabel(12.8097,8.87979)[bl]{$\tilde q_i^s$}
\FAVert(10.,14.5){0}
\FAVert(6.,5.5){0}
\FAVert(14.,5.5){0}
\FAVert(10.,11.5){0}

\FADiagram{}
\FAProp(0.,15.)(4.,10.)(0.,){/Straight}{1}
\FALabel(1.26965,12.0117)[tr]{$u$}
\FAProp(0.,5.)(4.,10.)(0.,){/Cycles}{0}
\FALabel(2.73035,7.01172)[tl]{$g$}
\FAProp(20.,15.)(16.,14.)(0.,){/Straight}{-1}
\FALabel(17.6241,15.5237)[b]{$u$}
\FAProp(20.,5.)(16.,6.)(0.,){/ScalarDash}{0}
\FALabel(17.6847,4.71881)[t]{$h^0$}
\FAProp(4.,10.)(9.5,10.)(0.,){/Straight}{1}
\FALabel(6.75,11.07)[b]{$u$}
\FAProp(16.,14.)(16.,6.)(0.,){/ScalarDash}{-1}
\FALabel(17.07,10.)[l]{$\tilde u^s$}
\FAProp(16.,14.)(9.5,10.)(0.,){/Straight}{0}
\FALabel(12.5487,12.6272)[br]{$\tilde g$}
\FAProp(16.,6.)(9.5,10.)(0.,){/ScalarDash}{-1}
\FALabel(12.4176,7.15993)[tr]{$\tilde u^t$}
\FAVert(4.,10.){0}
\FAVert(16.,14.){0}
\FAVert(16.,6.){0}
\FAVert(9.5,10.){0}

\FADiagram{}
\FAProp(0.,15.)(5.5,14.5)(0.,){/Straight}{1}
\FALabel(2.89033,15.8136)[b]{$u$}
\FAProp(0.,5.)(5.5,5.5)(0.,){/Cycles}{0}
\FALabel(2.89033,4.18637)[t]{$g$}
\FAProp(20.,15.)(14.5,14.5)(0.,){/Straight}{-1}
\FALabel(17.1097,15.8136)[b]{$u$}
\FAProp(20.,5.)(14.5,5.5)(0.,){/ScalarDash}{0}
\FALabel(17.1323,4.43534)[t]{$h^0$}
\FAProp(5.5,14.5)(5.5,5.5)(0.,){/ScalarDash}{1}
\FALabel(4.43,10.)[r]{$\tilde u^s$}
\FAProp(5.5,14.5)(14.5,14.5)(0.,){/Straight}{0}
\FALabel(10.,15.32)[b]{$\tilde g$}
\FAProp(5.5,5.5)(14.5,5.5)(0.,){/ScalarDash}{1}
\FALabel(10.,4.43)[t]{$\tilde u^s$}
\FAProp(14.5,14.5)(14.5,5.5)(0.,){/ScalarDash}{-1}
\FALabel(15.57,10.)[l]{$\tilde u^t$}
\FAVert(5.5,14.5){0}
\FAVert(5.5,5.5){0}
\FAVert(14.5,14.5){0}
\FAVert(14.5,5.5){0}

\end{feynartspicture}
\end{footnotesize} 

\caption{\label{ughu}
Typical Feynman graphs for the process $u g \to u h^0$ in
leading order. 
For the scattering of the other quarks ($d,c,s$) the graphs look similar.
}
\end{center}
\end{figure}

\end{document}